\documentclass[sigconf]{acmart}

\usepackage{multirow}
\usepackage{longtable}
\usepackage{xspace}
\usepackage{wrapfig}
\usepackage{graphicx,subcaption,lipsum}

\usepackage{enumitem}

  {\list{}{\selectfont\itshape\leftmargin=0.25in\rightmargin=0.25in}\item[]}%
  {\endlist}

\AtBeginDocument{%
  \providecommand\BibTeX{{%
    \normalfont B\kern-0.5em{\scshape i\kern-0.25em b}\kern-0.8em\TeX}}}

\newcommand{\rev}[1]{#1}
\newcommand{\nathalie}[1]{}
\newcommand{\ken}[1]{}
\newcommand{\michel}[1]{}
\newcommand{\nic}[1]{}
\newcommand{\hugo}[1]{}
\newcommand{\fred}[1]{}
\newcommand{\dave}[1]{}

\copyrightyear{2025} 
\acmYear{2025} 
\setcopyright{cc}
 \setcctype{by}
 \acmConference[CHI '25]{CHI Conference on Human Factors in Computing
 Systems}{April 26-May 1, 2025}{Yokohama, Japan}
 \acmBooktitle{CHI Conference on Human Factors in Computing Systems (CHI
 '25), April 26-May 1, 2025, Yokohama,
 Japan}
 \acmDOI{10.1145/3706598.3714259}
 \acmISBN{979-8-4007-1394-1/25/04}

\begin{document}

\title[AI-Instruments]{~\textit{AI-Instruments}: Embodying Prompts as Instruments to Abstract \& Reflect Graphical Interface Commands as General-Purpose Tools}

\author{Nathalie Riche}
\affiliation{
  \institution{Microsoft Research}
  \city{Redmond}
  \state{WA}
  \country{USA} 
}
\email{nath@microsoft.com}

\author{Anna Offenwanger}
\affiliation{
  \institution{Microsoft Research }
  \city{Redmond}
  \state{WA}
  \country{USA} 
}
\affiliation{
  \institution{Université Paris-Saclay, CNRS, Inria}
  \city{Orsay}
    \country{France} 
}
\email{anna.offenwanger@gmail.com}

\author{Frederic Gmeiner}
\affiliation{
  \institution{Microsoft Research}
    \city{Redmond}
  \state{WA}
  \country{USA} 
  }
  \affiliation{
  \institution{Carnegie Mellon University}
      \city{Pittsburgh}
  \state{PA}
  \country{USA} 
}
\email{gmeiner@cmu.edu}

\author{David Brown}
\affiliation{
  \institution{Microsoft Research}
  \city{Redmond}
  \state{WA}
  \country{USA} 
}
\email{dabrown@microsoft.com}

\author{Hugo Romat}
\affiliation{
  \institution{Microsoft}
  \city{Redmond}
  \state{WA}
  \country{USA} 
}
\email{romathugo@microsoft.com}

\author{Michel Pahud}
\affiliation{
  \institution{Microsoft Research}
  \city{Redmond}
  \state{WA}
  \country{USA} 
}
\email{mpahud@microsoft.com}

\author{Nicolai Marquardt}
\affiliation{
  \institution{Microsoft Research}
  \city{Redmond}
  \state{WA}
  \country{USA} 
}
\email{nicmarquardt@microsoft.com}

\author{Kori Inkpen}
\affiliation{
  \institution{Microsoft Research}
  \city{Redmond}
  \state{WA}
  \country{USA} 
}
\email{kori@microsoft.com}

\author{Ken Hinckley}
\affiliation{
  \institution{Microsoft Research}
  \city{Redmond}
  \state{WA}
  \country{USA} 
}
\email{kenneth.p.hinckley@gmail.com}

\renewcommand{\shortauthors}{Riche et al.}

\begin{abstract}

Chat-based prompts respond with verbose linear-sequential texts, making it difficult to explore and refine ambiguous intents, back up and reinterpret, or shift directions in creative AI-assisted design work. 
~\textit{AI-Instruments} instead embody “prompts” as interface objects via three key principles: (1)~\textit{Reification} of user-intent as reusable direct-manipulation instruments; (2)~\textit{Reflection} of multiple interpretations of ambiguous user-intents (\textit{Reflection-in-intent}) as well as the range of AI-model responses (\textit{Reflection-in-response}) to inform design "moves" towards a desired result; and (3)~\textit{Grounding} to instantiate an instrument from an example, result, or extrapolation directly from another instrument. Further, AI-Instruments leverage LLM’s to suggest, vary, and refine new instruments, enabling a system that goes beyond hard-coded functionality by generating its own instrumental controls from content. 
We demonstrate four technology probes\rev{, applied to image generation,} and qualitative insights from twelve participants, showing how AI-Instruments address challenges of intent formulation, steering via direct manipulation, and non-linear iterative workflows to reflect and resolve ambiguous intents.

\end{abstract}



\begin{teaserfigure}
  \includegraphics[width=\textwidth]{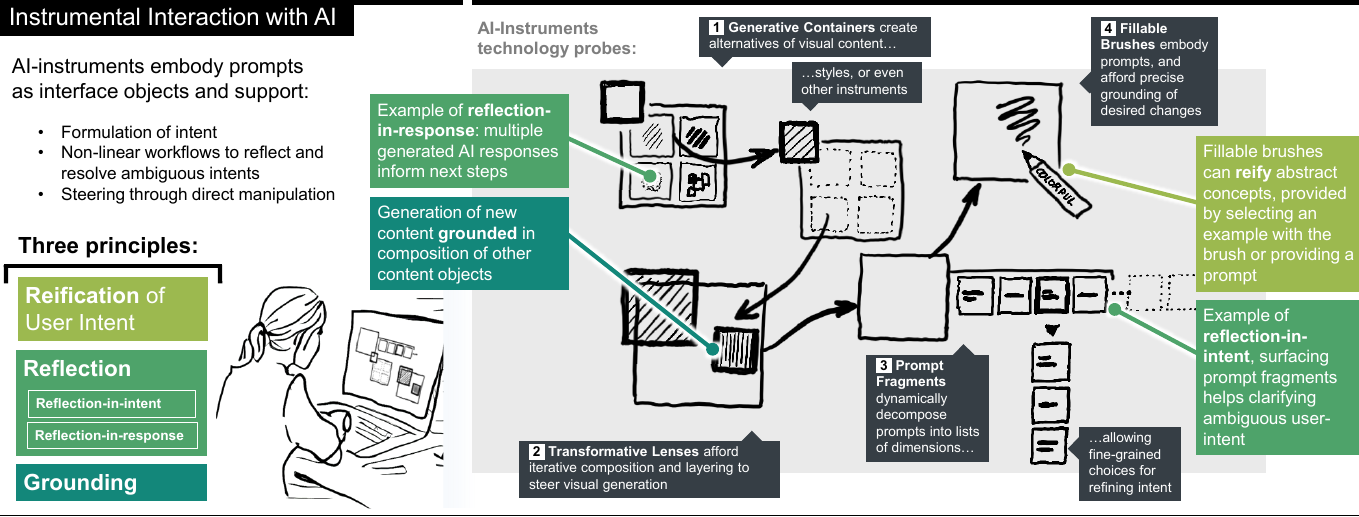}
  \caption{AI-Instruments embody prompts as interface objects, informed by three principles: reification of user intent, reflection, and grounding (left). Visual overview of four technology probes of AI-Instruments -- generative containers, transformative lenses, prompt fragments, and fillable brushes (right).}
  \Description{Visual Abstract of the paper. Visual illustration of a person using AI-Instruments on the left, then a box including the three principles of AI-Instruments in the middle, and sketched illustrations of AI-Instruments on the right (using GUI elements and wireframes). The figure explains that AI-Instruments embody prompts as interface objects, informed by three principles (left): reification of user intent, reflection, and grounding. (right) Visual overview of four technology probes of AI-Instruments -- generative containers, transformative lenses, prompt fragments, and fillable brushes.}
  \label{fig:teaser}
\end{teaserfigure}

\maketitle

\section{Introduction}

Despite the immense promise of generative AI, it remains challenging for people to express and refine their true intents via multiple rounds of textual chat-prompts~\cite{mahdavi_goloujeh_is_2024, zamfirescu-pereira_why_2023}, as well as to pursue multiple-alternative paths forward within a linear conversational metaphor. 
Users face numerous difficulties (e.g.,~\cite{subramonyam_bridging_2024}): articulating their intent in a few words of written text (intent formulation); correctly expressing sufficient detail to express, refine, and re-formulate their true intent (intent disambiguation); iterating over the model's response to approach a  desired outcome (steering); and navigating higher-order challenges of interaction with AI, such as discovering what one can do with AI---or even what one "actually" wants to achieve (intent resolution)---within the linear sequential limitations of chat-based exchanges (interaction workflow). 

Although existing work in human-computer interaction and artificial intelligence (HCI+AI) addresses some aspects of these challenges in piecemeal fashion through novel interaction techniques with generative AI (e.g.~\cite{chung_promptpaint_2023, masson_directgpt_2024}), we argue here that ~\textit{AI-Instruments} offer a novel approach to gain traction on many aspects of these challenges by appropriating and re-casting the principles of ~\textit{instrumental interaction} ~\cite{beaudouin2000instrumental, beaudouin2000reification} to the modern context of generative HCI+AI user experiences. \rev{Quoting Beaudoin-Lafon et al.~\cite{beaudouin2021generative}, the value of proposing such  interaction model is to \textit{"change-oriented perspective by providing HCI researchers with conceptual tools for analyzing technologies in use or exploring novel future solutions"}. Triangulating theory, artifact, and empirical evaluation has strong benefits for advancing HCI research~\cite{mackay1997hci}.}

Instrumental interaction offers a particularly compelling concept from the HCI literature to revisit in the context of generative AI because it offers principled interaction dynamics about how software functionalities (\textit{"commands"}) combine with content (the \textit{"objects"} those commands act upon). While in the past these dynamics had to be hand-designed and hand-coded for specific object types and application settings, the advent of generative AI makes it plausible that the polymorphic nature of high-level commands and flexible content representations will unleash exciting new possibilities for HCI+AI graphical user interfaces. 

In particular, our approach \textit{embodies AI prompts as graphical interface objects} and adapts the instrumental interaction model for Generative AI by considering the following three principles: 

\begin{enumerate}[label={(\arabic*)}]
    \item [(1)] \textit{\textbf{Reification of user intent}} into instruments: turning user-intent from varied abstractions and granularity levels into one or more reusable graphical interface object(s);
    \item  [(2)] \textit{\textbf{Reflection}}: the consideration of multiple alternatives that reflect ~\cite{DesignReflectiveConversation1992} both ambiguous intents as expressed by the user (\textit{reflection-in-intent}), and ambiguous interpretation of AI responses (\textit{reflection-in-response}), to steer content generation towards a satisfactory result; and finally
    \item [(3)] \textit{\textbf{Grounding}}: instantiating an instrument from a specific scope of selected content, from an example result, or even from another instrument.
\end{enumerate}

Via a technology probe~\cite{TechnologyProbesCHI2003} that implements four complementary examples of AI-instruments, we illustrate how they can ameliorate many design challenges plaguing today’s linear-chat-based AI interfaces: intent formulation, prompt engineering, direct manipulation and steering, non-linear iterative workflows, and intent resolution. We also present initial reactions of 12 participants who tried our AI-instruments, yielding qualitative insights on the value and limitations of our AI-instruments interaction model, as compared to conversational prompting.

Designed through the lens of the three principles, \rev{we built a set of technology probes focused on image generation. The goal of these four exemplar AI-Instruments is to demonstrate the new interaction capabilities and affordances of our model:} (1) \textit{Fragments} decompose gen-AI prompts into reified reconfigurable objects, affording reflection-on-intent on the latent prompt structure, and grounding generation by dragging fragments from one object to another. (2) \textit{Transformative Lenses} generate new content grounded in one or more content elements, which allows flexible recomposition of scenes and (if desired) continuous updates of the result. (3) \textit{Generative Containers} create multiple alternatives of images, text, and even instruments or fragments. 
(4) \textit{Fillable Brushes} encapsulate a prompt, filled by selecting example content with the brush (or by directly typing the prompt for a new action). Using the instruments in synergy---where outputs from one instrument form input for the next, or even using instruments to create new meta-instruments---affords expressive degrees-of-freedom for fine-grained steering of generative AI.

In summary, our high-level contributions include the following:
\begin{itemize}
    \item Extend the classic instrumental interaction model~\cite{beaudouin2000instrumental} to generative AI, emphasizing three driving principles: reification of user intent, reflection, and grounding;
    \item Demonstrate four AI-instruments via technology probes, showing how these driving principles manifest in their design and implementation;
    \item Provide initial reactions from 12 users when shifting from a linear-chat interaction paradigm to direct manipulation through AI-instruments, showing that it can address a number of human-AI interaction challenges.
\end{itemize}

In the following sections we discuss related techniques across the HCI, Human-AI interaction, and design literature. This is followed by an Example Walkthrough of our AI-instruments, a wider discussion of Instrumental Interaction with AI, and further details of our Four Exemplar AI-Instruments: Fragments, Transformative Lenses, Generative Containers, and Fillable Brushes. We then present a qualitative Study of these AI-Instruments in comparison with textual prompting, and finally close with a Discussion and Future Work.

\section{Related Work}

We first discuss the state of human-AI interaction, articulating it around five core challenges. Then, we motivate the need for a more general interaction model and point to research in design and creativity grounding two principles we introduce.

\subsection{Human-AI Interaction}
\label{sec:challenges}
Recent work explores the difficulties users face when interacting with generative AI via prompting~\cite{zamfirescu-pereira_why_2023, subramonyam_bridging_2024, mahdavi_goloujeh_is_2024, peng_designprompt}. Earlier research identified barriers that arise (for example) in end-user programming~\cite{ko2004six} and, more generally, bridging the gulf of execution and the gulf of evaluation ~\cite{norman1986cognitive}. We organize emerging research for interaction with generative AI under five core Challenges (C1-C5) faced by users, and discuss later in the paper how our interaction model addresses each.

\vspace{5pt}\textit{\textbf{(C1) Intent formulation}} via prompting, solely using natural language, can be challenging when the outcome is hard to describe in words. 
Users may lack the vocabulary to describe visual styles, or the high-level impressions they seek to achieve. Researchers studied thousands of prompts to generate images~\cite{liu_design_2022} to develop guidelines for prompting and parameter selections. They coupled prompting with images to offer richer multimodal intent formulation. PromptCharm~\cite{wang_promptcharm_2024} leveraged a large image database to help users find the right style of images and incorporated interactive techniques -- such as linking a prompt fragment to the corresponding part of the generated image, to provided richer solutions for users to formulate their intent. Similarly, DesignPrompt~\cite{peng_designprompt} affords expressive multi-modal prompt construction. Such research seeking to expand the modalities we have to communicate with models beyond text input is particularly important~\cite{liu_beyond_2023} for multimodal outputs such as generated videos~\cite{villegas2022phenaki}, 3D objects~\cite{poole2022dreamfusion}, and virtual worlds~\cite{rosenberg_drawtalking_2024}. 

\vspace{5pt}\textit{\textbf{(C2) Intent disambiguation}} is the skill of describing one's intent with enough specific detail for AI to produce the intended result. Much past work on \textit{prompt engineering} across several fields of research tackles this challenge, with research probes of this issue ~\cite{zamfirescu-pereira_why_2023} suggesting templates and guidelines~\cite{bozkurt_tell_2024} for users to provide the information they might have difficulty thinking about upfront. Beyond prompt engineering, the HCI community explores different representations to facilitate communicating context to the system. For example, Graphologue~\cite{jiang_graphologue_2023} represents a prompt as an interactive node-link diagram that users can expand and complete to incrementally add context to their intent. Such work also addresses the ambiguity of natural language by enabling users to unpack certain parts of their intent and disambiguate them by adding more information. \rev{Promptify~\cite{brade_promptify_2023} organizes generated content on a canvas based on a person's preferences and suggests alternatives -- leading to an iterative loop with the user refining, selecting, and discarding alternatives of prompts and content.}

\vspace{5pt}\textit{\textbf{(C3) Intent resolution}} is the challenge users confront to determine what outcomes may or may not match their original intent. Difficulties here may stem from an ambiguity of intent in the users' mind (e.g. a user might realize \textit{"I am not even sure what exact outcome I want"}). This problem, as a well-known attribute of challenging creative design work ~\cite{DesignReflectiveConversation1992, ReflectivePhysicalPrototyping2006, BuxtonSketchingUserExperiences2007}, is certainly not unique to AI but may be exacerbated by the relative novelty, black-box nature, and rapidly accelerating capabilities of modern AI models~\cite{bubeck2023sparks}. 
However, further difficulties may arise from people's lack of knowledge of what an AI model can or cannot do. 
Here, approaches from graphic design may help users explore possible outcomes, such as CreativeConnect~\cite{choi_creativeconnect_2024}, which extracts keywords, and text descriptions from a set of reference images and facilitate recombination and reuse. Other work shows the possibilities of what users can ask via prompt-space exploration~\cite{almeda_prompting_2024}, or through interfaces that reveal what results a user can generate~\cite{suh_luminate_2024}.

\vspace{5pt}\textit{\textbf{(C4) Steering}} the result of generative AI to get closer to either what the user initially imagined or to an unforeseen result assessed as satisfactory is a fundamental human-AI interaction mechanism. The topic has been studied for multiple decades in multiple field and referred to as human-in-the-loop~\cite{wu2022survey} and mixed-initiative interfaces~\cite{horvitz1999principles}. Within the context of generative AI, research on the topic has centered on human-AI co-creation~\cite{davis2015enactive}. Researchers developed human-AI co-creation interfaces for specific activities such as drawing~\cite{oh_i_2018}, crafting images~\cite{chung_promptpaint_2023} and writing stories~\cite{chung_talebrush_2022, zhang_storydrawer_2022}. These interfaces either surface generative AI capabilities as graphical interface elements such as a button to generate a character for a story~\cite{zhang_storydrawer_2022}, or propose custom graphical widgets to specify constraints or parameters of the content to be generated by the model such as an interactive line chart depicting the narrative arc of the story~\cite{chung_talebrush_2022}.

Recent research has begun to explore more generic interaction solutions to the prompting chat-based experiences incorporated in most mainstream products today. Steering content generation in conversational prompting amounts to a linear trial-and-error process, in which users type a prompt, and then evaluate its result. They then must either rerun the same prompt to get a new result (since generative AI is non-deterministic); or edit the prompt to get an iteration over the prior result. By building upon principles of \textit{direct manipulation}, DirectGPT~\cite{masson_directgpt_2024} offers an early glimpse of an alternative interaction human-AI co-creation paradigm based on the principles of direct manipulation and surfaced to users with graphical widgets (e.g. buttons) that might generalize to a wider range of outputs and applications. Our research extends this ambition via instrumental interaction~\cite{beaudouin2000instrumental}, yielding a novel interaction model that can provide the community with both evaluative (assessing novel interaction techniques) and generative (inspiring the design of novel interaction techniques) power. 

\vspace{5pt}\textit{\textbf{(C5) Interaction workflow}} models based on conversation with generative AI are inherently linear.  Research started to investigate non-linear interaction workflows with generative AI. In particular, DeckFlow~\cite{croisdale_deckflow_2023} relies on mood board type interaction and also breaks the silo of different models. Sensecape~\cite{suh_sensecape_2023} and Graphologue~\cite{jiang_graphologue_2023} leverage additional non-linear metaphors to enable people to perform non-linear interactions with AI.
\rev{These systems focus on a specific metaphor for conversation with LLMs, laying out prompts and responses as a graph in a canvas.}
\rev{Graph structures can also function as an intermediary representation facilitating prompt steering, by breaking down text prompts into hierarchical structures of granular elements \cite{xcreation}. Similarly, tree structures enable traversing alternative representations of generated content, where sub-nodes represent distinct visual aspects across the latent space \cite{conceptdecomposition}. Our intent is to identify general principles that afford direct manipulation for decomposing and (re)composing  objects at multiple levels of granularity for different tasks and contexts.}

The interaction model we propose provides a generic solution to address these 5 challenges by building upon the instrumental interaction model and apply it to the context of building interactive applications leveraging generative AI capabilities.

\subsection{Interaction Models in the Era of AI}

Despite tremendous advances in technology and the promise of Artificial General Intelligence~\cite{bubeck2023sparks}, mainstreams interfaces today feature a chat-based interface with AI reminiscent of command-line human-computer interaction paradigm of the 1960s. While the use of natural language does remove barriers of adoption for the general public, many of the limitations of communicating instructions in a linear and sequential manner by typing, later addressed by Graphical User Interfaces, pertain.

Over the years, the HCI community has produced knowledge on human-computer interaction~\cite{hornbaek2017interaction}, devised principles and theories for improving interaction~\cite{hutchins1985direct,  gibson1977theory}, and proposed multiple interaction models~\cite{beaudouin2000instrumental,jacob2008reality} for building the next generation of interfaces. These models are generally grounded in the emerging interfaces and techniques of the time, surfacing key principles governing them and desirable properties when humans interact with them. The goal of these models is to inform and assess the design of the next generation of interfaces. Our work has the same ambition: \textit{informing and guiding the design of interfaces leveraging generative AI}. While numerous recent work centered on advancing specific use cases and application areas -- seeking to identify and leverage the value of generative AI -- few researchers relate to existing theory and models, or proposing \textit{new theories and models} in this era of AI. \rev{Perhaps the closest effort is the Cells, Generators, and Lenses model proposed by Kim et al.~\cite{kim2023cells} which proposes a design framework for helping designers identify and reflect on basic building blocks needed for interfaces leveraging AI}. \rev{Our research is complementary to this effort, seeking to identify interaction principles that afford direct manipulation of these building blocks}.

Our work seeks to build upon and extend the instrumental interaction model to the design of generative AI interfaces. The instrumental interaction model~\cite{beaudouin2000instrumental} directly builds upon direct manipulation and generalizes the use of instruments to mediate between user and objects of interests (e.g. content). It describes a large range of interaction techniques that were not captured in WIMP and direct manipulation such as lenses or tangible interactions. \rev{Recent work attempted to leverage the instrumental interaction model to design novel interactions with AI. For example, Yen and Zhao~\cite{yen2024memolet} used reification to turn prior conversations with AI into graphical objects or Memolets, that users can interact with.
Our work propose a more general adaptation of this model to content generation with AI and expands it with additional principles of reflection and grounding. The resulting model we propose falls into generative theories of interaction~\cite{beaudouin2021generative}, aiming at inspiring and informing the design of novel techniques.}

\subsection{Content Generation and Creativity Support}

Several key insights from the design and creativity support literature~\cite{shneiderman_creativity_2007} motivate our principles of reflection and grounding.

Design and creativity processes embrace ambiguity of low-fidelity prototypes~\cite{BuxtonSketchingUserExperiences2007} and rapid cycles of idea generation and evaluations~\cite{TEAMSTORM2007, TerryCreativeNeedsUIDesign2002, ReflectivePhysicalPrototyping2006} to enable people to explore many design alternatives, reflect on their possibilities through the action of sketching and building, and iterate on the most promising ones ~\cite{DesignReflectiveConversation1992}. Researchers have also described these processes as sequences of divergent thinking followed by convergent thinking~\cite{farooq2005supporting}. As fundamental working-patterns that people exhibit in challenging content creation and design tasks, there is good reason to believe such processes should persist and be supported by tools for creative content generation with generative AI. Such tools should help users rapidly investigate alternatives ideas in fluid, non-linear manner (e.g. exploration) and support the rapid iteration of the most promising content (e.g. steering). Direct manipulation and instrumental interaction models offer a compelling point of departure, affording chunking and phrasing~\cite{buxton1995chunking} of complex generative-AI interactions for exploration and steering.

As hinted above, our principle of reflection builds on Schön's notion of \textit{reflection-in-action} 
~\cite{DesignReflectiveConversation1992}---where the externalized materials of design "speak to" the designer to help them reflect-\textit{on}-action as to the next design "move" to make within an ambiguous space of many possible ideas. 
In the context of generative AI, this principle of ~\textit{reflection} conveys the notion that instruments should reflect the design space of user intent---as well as the wide potential space of generated results---to help users make informed decisions ("moves") as they iterate towards a desired (AI-assisted) outcome. A related concept to reflection and idea incubation is the process of gathering inspirational materials in moodboards~\cite{cassidy2008mood, BuxtonSketchingUserExperiences2007}. Such design practices help identify concepts and themes, especially when these are hard to articulate in words, or isolate from one another other~\cite{freeman2017creativity}. We refer to this activity in our principle of grounding, to convey the idea that instruments can extract specific aspects from a set of materials, and to then apply them to different content.

\rev{An aspiration for an interaction model geared on content generation is to afford power to their users~\cite{li2023beyond}. In particular, vertical movement (moving up and down the abstraction ladder) afforded by natural language input of LLMs; and horizontal movement (composing tools and workflows) afforded by combining instruments together offer promising avenues for AI-instruments.}

\begin{figure*}
    \centering
    \includegraphics[width=1.0\linewidth]{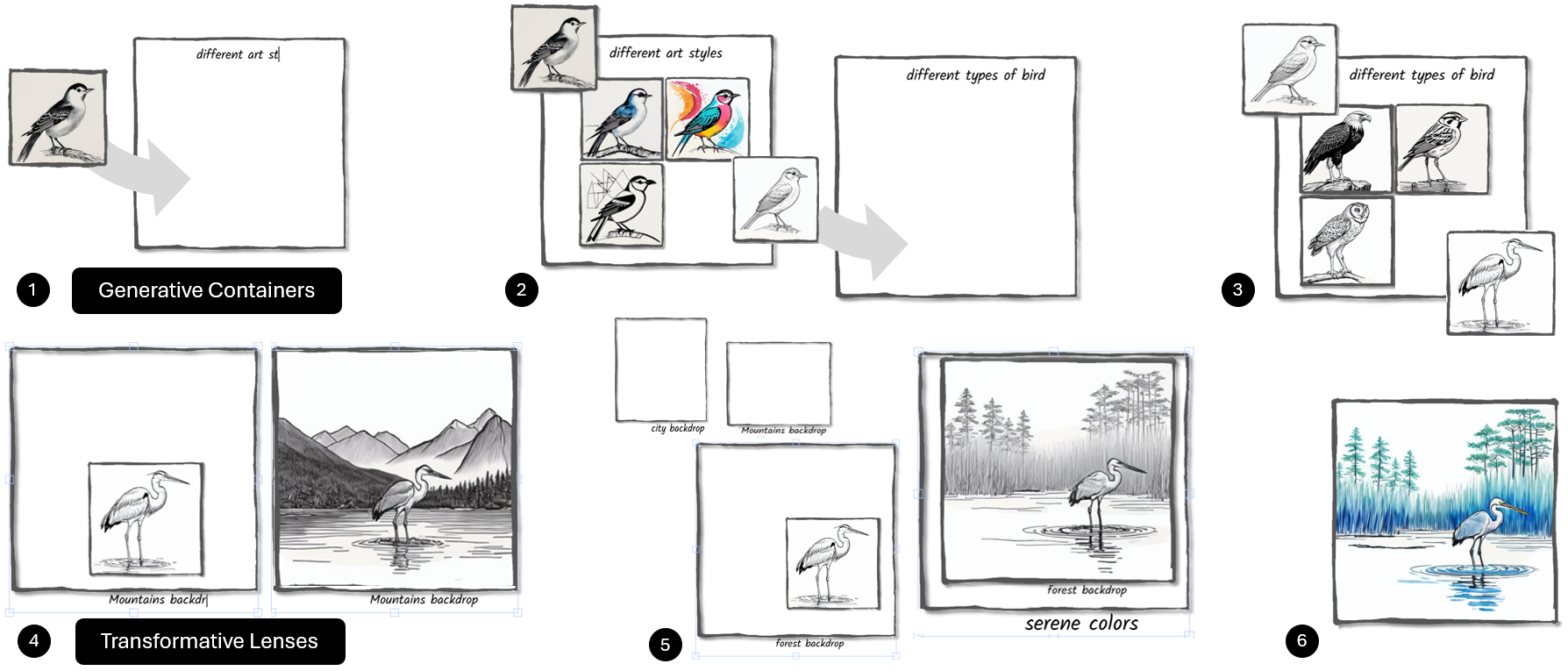}
      \caption{\rev{Sequence of interactions to explore ideas with generative containers and lens probes: When dragging an image into a container (1), variations are created based on \textit{style} (2). When selecting one of these images and dragging it into another container with the prompt "different types of bird", variations of different kinds of birds are generated in a consistent art style (3). A transformative lens around one of the earlier images generates a landscape around the bird through inpainting (4), and allows more complex composition of content (5, 6).}}
      \Description{Screenshots of the sequence of interactions to explore ideas with generative containers and lenses probes. The figure shows a number of wireframes that illustrate the interaction sequence. 1. dragging an image into a container, 2. variations are created based on style and one image of that one is dragged into another container. 3. Variations of different kinds of birds are generated in a consistent art style. 4. A transformative lens adds a frame around one of the earlier images and shows how the inpaint method generates a landscape around the bird. 5. a more complex images is composed with the help of a transformative lens. 6. a final colorful image of the bird is shown as the final result of the pipeline.}
    \label{fig:walkthrough1}
\end{figure*}
\begin{figure*}
    \centering
   \includegraphics[width=1.0\linewidth]{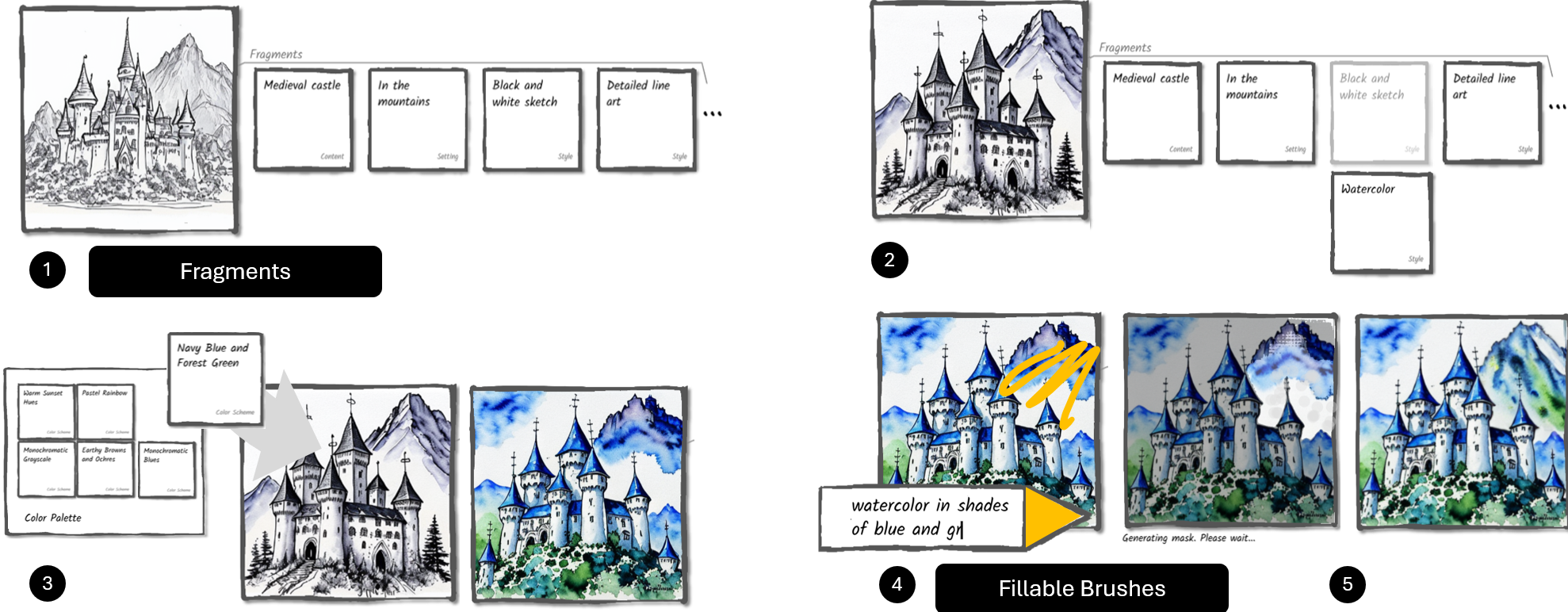}
      \caption{\rev{Sequence of interactions to steer image generation with fragments and brushes probes: Prompt fragments are generated for an existing image and show dimensions of the image to manipulate (1). A person can modify any of these fragments and a new image is generated (2). Containers can generate variations of fragments, which are then used to modify the image (3). Fillable Brushes (pen-like instruments) are used to modify the image of a castle, changing the art rendering style and color where the brush painted over the image, based on the prompt that was 'filled' into the pen (4, 5).}}
      \Description{Screenshots of the sequence of interactions to steer image generation with fragments and brushes probes. Different software wireframes show the interactions with the AI-Instruments. 1. an image is shown, with a number of text boxes on the right side which represent the 'fragments' the AI extracted. 2. A person then modifies one of these fragments and a different image is generated. 3. a container generates variations of fragments, which are then used to modify the image. 4. Fillable Brushes (pen-like instruments) are used to modify the image of a castle, changing the art rendering style and color where the brush painted over the image.}
    \label{fig:walkthrough2}
\end{figure*}

\section{Example Walkthrough}

Let us take the example of Emma, who is seeking to illustrate a social media post to express the serenity she feels when she spends time outdoors (Figure~\ref{fig:walkthrough1}). She starts from leveraging generative AI to generate a bird. The art style is not quite satisfying but she is not sure what the model is capable of. She selects a generative container from a panel of AI-instruments available to her (\rev{Figure~\ref{fig:walkthrough1}.}1 and 2) and explores different art styles. She finds a simple drawing style she likes, and creates a second generative container to explore other types of birds in the same style (\rev{Figure~\ref{fig:walkthrough1}.}3). She settles on a heron, and moves on composing a more interesting illustration. Since she has an idea of the general composition she wants, she opts for a transformative lens, a second AI-instrument enabling her to position her central character, the bird, in the frame (\rev{Figure~\ref{fig:walkthrough1}.}4). She creates multiple lenses to try multiple backdrops and settles on the forest one (\rev{Figure~\ref{fig:walkthrough1}.}5). As lenses can be layered, she creates a color style one, and layers it on top of the forest backdrop, resulting in an illustration she finds suitable for her post (\rev{Figure~\ref{fig:walkthrough1}.}6).

Two days later, Emma seeks to illustrate her school presentation on medieval castles (Figure~\ref{fig:walkthrough2}). She starts from a drawing generated by AI. She taps-and-holds to expand the fragments the model used to generate the image (\rev{Figure~\ref{fig:walkthrough2}.}1). By tapping on different fragments, she gets to try different variations, such as redrawing the castle as watercolor style (\rev{Figure~\ref{fig:walkthrough2}.}2). As she wants to add color to the illustration, Emma retrieves the palette where she saved multiple fragments related to colors she thought worked great in the past (\rev{Figure~\ref{fig:walkthrough2}.}3), and drags one onto the image. She does like the colors but notices a large white space in the back. She selects a fillable brush, \rev{an AI-instrument that lets} her directly scrub over the specific portions of the image that she wants to revise or refine (\rev{Figure~\ref{fig:walkthrough2}.}4). After she types the outcome she wants and brushes the region, the system generates a mask and applies changes locally (\rev{Figure~\ref{fig:walkthrough2}.}5). Emma is now satisfied with her illustration.

\section{Instrumental Interaction with AI}

Beaudoin-Lafon defines instruments as: \textit{"a mediator or two-way transducer between the user and domain objects."} we expand this definition to AI-instruments: \textit{"an AI-powered mediator or two-way transducer between the user and domain objects."} We describe below the three principles of our proposed model revision: reification of user intent, reflection and grounding. Note that these principles are tightly interconnected and, while differing in certain aspects from the original model also share a lot of similarities. We discuss differences in more depth in Discussion.

\subsection{Reification of User Intent}

Most pre-AI interfaces offer a finite set of functionalities, established at their design by software architects and developer. User experience designers craft a set of graphical interface components and interactions for each functionality to enable users to invoke a finite set of commands through this GUI. Today, LLMs can interpret requests from users in natural language and turn them into the execution of a specific command, or a sequence of commands, unbounding functionalities from a limited set of GUI components. With this major shift in interface design, we propose the reification of \textbf{user intent}, rather than \textbf{commands}.

\begin{figure}[t]
    \centering
\includegraphics[width=.48\textwidth]{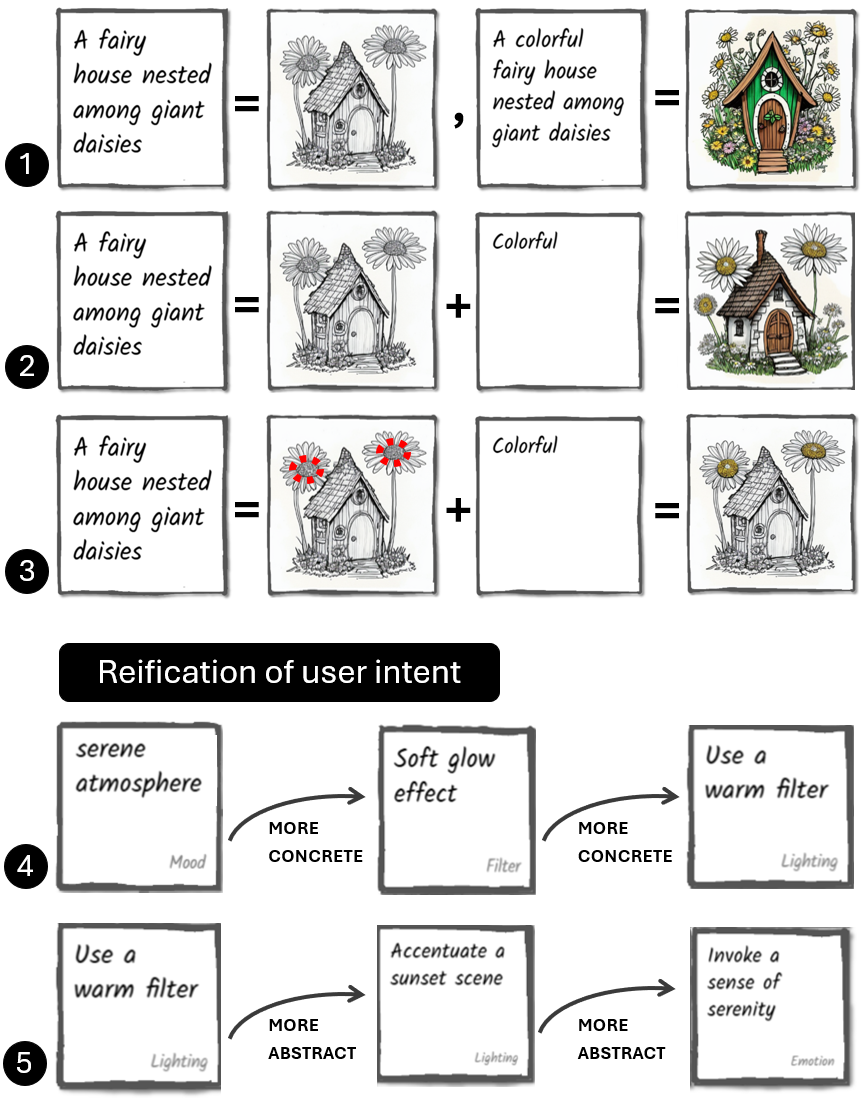}
\caption{In the chat-based interaction model, interactions consists of a linear sequence of input+output pairs and steering is done by modifying the input (1). Reification enables articulating interactions into phrases for example by reusing the output of the prior input (2). It also affords direct manipulation techniques such as for lasso selection (in red) to specify the scope of the input (3). Reification of user intent enables users to reflect on their intent and navigate dimensions such as its degree of abstraction, using other instruments to make it more concrete (4) or abstract (5) for example.}
\Description{Screenshots of wireframes and examples illustrating the 'reification of user intent'. In the classic chat-based interaction model, interactions consists of a linear sequence of input+output pairs and steering is done by modifying the input (1). Reification enables articulating interactions into phrases for example by reusing the output of the prior input (2). It also affords direct manipulation techniques such as for lasso selection (in red) to specify the scope of the input (3). Reification of user intent also enables users to reflect on their intent and navigate dimensions such as its degree of abstraction, using other instruments to make it more concrete (4) or abstract (5) for example.}
    \label{fig:reification}
\end{figure}

Reification turns both input and output of generative AI into graphical elements that can be directly manipulated and thus reused by users. In contrast to chat-based interfaces consisting of sequences of [input+output] in \rev{which users can require} to rephrase the input to iterate, reifying input and output enables users to articulate phrases of interaction~\cite{buxton1995chunking} and afford direct manipulation techniques such as lasso selections to specify scopes of intent (Figure~\ref{fig:reification} (1-3)).
In section~\ref{sec:Ai-instruments}, we demonstrate how this instrumental model can leverage the full range of direct manipulation techniques the community developed such as magic lenses~\cite{bier2023toolglass} and attribute objects~\cite{xia2016object}, turning them into AI-instruments encapsulating user intent.

A key capability of generative AI models is their inherent ability to deal with the \textbf{degree of abstraction} of user intent. It offers unparalleled flexibility as users can express high-level or low-level intent. Examples in the literature leverage the high degree of abstraction for content generation. For example, Talebrush~\cite{chung_talebrush_2022} enables users to control the narrative arc of a story (where tension is in the story), which has many implications on the writing itself from adding or sequencing events differently in the story to subtly rewording the language.  Expressing high-level intents is a powerful ability, enabling people to shape content 
in ways that potentially lead to serendipitous discovery of alternative (potentially better) results. However, users face multiple challenges when results are unsatisfactory, understanding how to resolve ambiguity of their intent (C2) or thinking more crisply of the desired outcome (C3). These challenges often require users to lower the degree of abstraction of their intent. On the contrary, expressing intents with a low degree of abstraction lowers the chance to make serendipitous discoveries, as well as get into a class of unwanted model results, making it frustrating for users to steer content generation towards more major changes (C4) or conduct exploratory workflows (C5). These challenges often require users to increase the degree of abstraction of their intent. Figuring out how to navigate degrees of abstraction is a challenge in itself. Users may struggle turning an idea into a set of concrete changes or, conversely, articulate the overarching goal motivating specific changes. Users can leverage AI-instruments themselves to navigate the degree of abstraction of an intent, for example, by using a Generative Container to provide more concrete (resp. abstract) Fragments given one of high-degree (resp. low-degree) of abstraction (Figure~\ref{fig:reification} (4-5)).

\subsection{Reflection}

Seminal research demonstrated that it is critical to explore alternative designs early and throughout the whole process~\cite{tohidi2006getting, DesignGalleries1997}. It is particularly critical when working with AI because of its "black box" nature~\cite{bathaee2017artificial,hoffman2018explaining}, i.e. the inherent difficulty for users to understand how these models work, and the non-deterministic nature of their outputs.  To capture this aspect, we borrow the term \textbf{reflection} from the design literature and introduce it as a principle for AI-instruments. 

We define reflection as the ability to help users reflect on their possibly ambiguous intent (reflection-in-intent) as well as the ambiguous interpretation made by AI (reflection-in-response), and thus offer the ability to users to steer the content generation towards a satisfying result.

\begin{figure*}[t]
    \centering
\includegraphics[width=.8\textwidth]{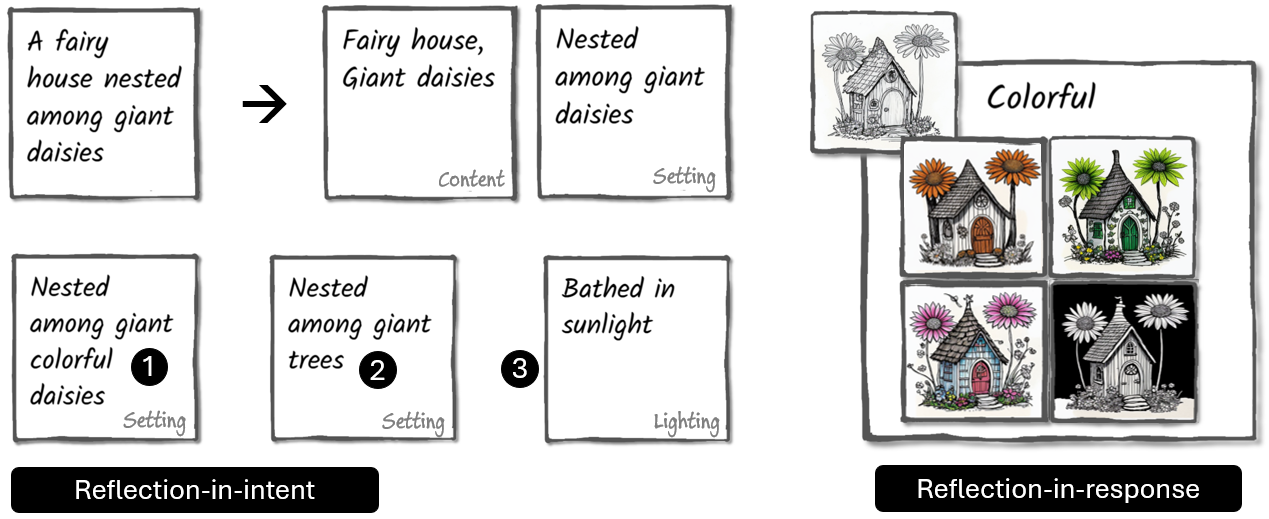}
\caption{Reflection-in-intent enables users to gain awareness of the possible  formulations of their intent while reflection-in-response enables users to assess the space of possibilities of the outputs generated by the model given an input. These aspects may help users address the challenges of intent disambiguation, resolution and steering.}
\Description{Screenshots of wireframes and examples illustrating the 'reflection-in-intent' and 'reflection-in-response'. Reflection-in-intent enables users to gain awareness of the possible  formulations of their intent while reflection-in-response enables users to assess the space of possibilities of the outputs generated by the model given an input. These aspects may help users address the challenges of intent disambiguation, resolution and steering.}
    \label{fig:reflection}
\end{figure*}

\textbf{Reflection-in-intent} is the ability of AI-instruments to surface multiple facets of their intent to users. For example,  fragmenting intent into pieces reveals a particular chunking~\cite{buxton1995chunking}. Working with fragments (Figure~\ref{fig:reflection}) may help users refine their intent (1), pivot on a specific aspect (2) or iterate by adding novel aspects (3).

\textbf{Reflection-in-response} is the ability of AI-instruments to offer multiple results of the content generation, while also helping people explore the space of possibilities (Figure~\ref{fig:reflection}), addressing (C3).Reflection-in-response can vary on the type and range of alternatives provided by employing diverse strategies: using model parameters such as its temperature, generating variations of the input, or asking the model to use different context of interpretation.

\begin{figure*}[t]
    \centering
\includegraphics[width=.8\textwidth]{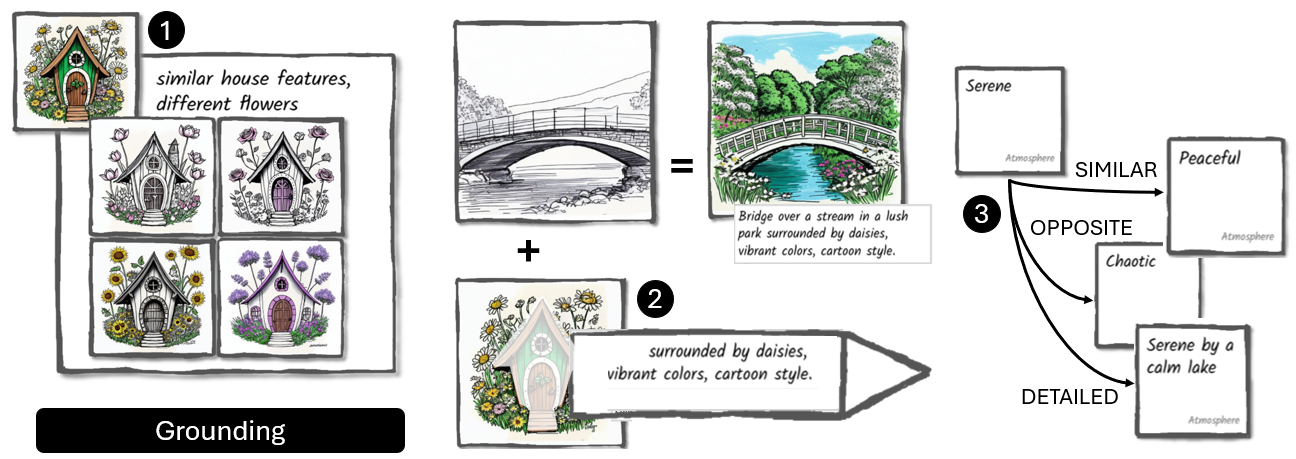}
\caption{Grounding an instrument such as a generative container with an example enables to refer to features to preserve or alter in simple worlds by leveraging AI segmentation (1). Grounding an instrument such as a fillable brush in a specific aspect of an example, for example by selecting a region and extracting its style (2), enables users to use and apply it to other inputs without the need to articulating it in words. The principle of grounding also applies to instruments themselves such as deriving fragments from an example one (3).}
\Description{Wireframes and screenshots used to explain the concept of grounding. Showing a number of images and different operations performed to modify the content. Grounding an instrument such as a generative container with an example enables to refer to features to preserve or alter in simple worlds by leveraging AI segmentation (1). Grounding an instrument such as a fillable brush in a specific aspect of an example, for example by selecting a region and extracting its style (2), enables users to use and apply it to other inputs without the need to articulating it in words. The principle of grounding also applies to instruments themselves such as deriving fragments from an example one (3).}
    \label{fig:grounding}
\end{figure*}

\subsection{Grounding}

The principle of grounding refers to the ability for users to ground instruments from examples of desired outcomes or other instruments. It may be difficult to find the right vocabulary to describe particular aspects of content, especially for images. Instruments leverage AI segmentation to (1) enable users to refer to elements of an example in generic terms,  and (2) extract specific aspects of the content (e.g. style) by selection, storing the result for later (Figure~\ref{fig:grounding}). This builds on the notion of \textit{Variations}, \textit{Parameter Spectrums}, and \textit{Side Views} \cite{TechnologyProbesCHI2003, TerrySideViews2002}, but in a way that leverages the principles of interactive instruments \cite{beaudouin2000instrumental, beaudouin2000reification} as well as the open-ended possibilities of generative AI via our novel AI-instruments, rather than as views or controls with fixed, hand-designed and hard-coded options. AI-instruments can also be grounded in other instruments, enabling exploration of the space of related instruments (Figure~\ref{fig:grounding} (3)).

\section{Examples of AI-Instruments}
\label{sec:Ai-instruments}
To assess the viability of our AI instrumental model, study its differences with existing GUIs and tease out its value compare to existing chat-based AI interaction, we built a technology probe~\cite{TechnologyProbesCHI2003} with four different instruments, grounded in the literature: Fragments, Generative Containers, Transformative Lenses and Fillable Brushes. 
We describe below how this set of instruments surface the principles of our AI-instrumental model, as well as offer complementary interaction capabilities and affordances.

\subsection{Fragments}
 Fragments build on the concept of \textit{Attribute Cards} introduced in Object Oriented Drawing~\cite{xia2016object}, as well as Side View's notions of \textit{Variations} and \textit{Parameter Spectrums} \cite{TerryCreativeNeedsUIDesign2002, TerrySideViews2002}, by using a large language model to extract multiple conceptual dimensions that may be plausibly implied by a prompt. 
 
\begin{figure*}[t]
    \includegraphics[width=0.88\textwidth]{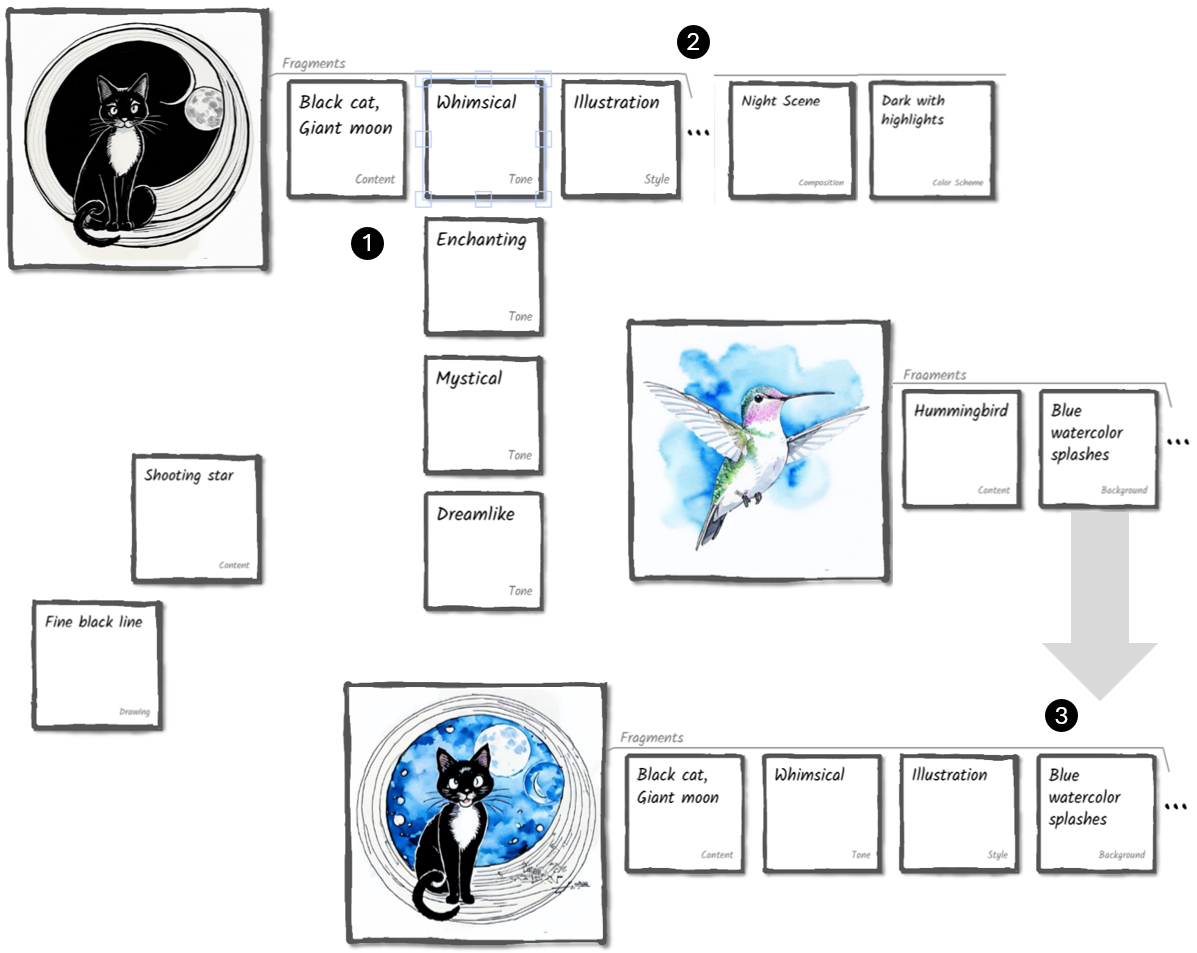}
    \caption{Users can expand Fragments with variations of parameter values (1) in vertical columns, or request more suggestions for dimension types (2) at the end of the row. Users can further reuse and transfer Fragments to other content via drag-and-drop(3).}
    \Description{Several wireframe boxes with images and text explain the concept of Fragments. Users can expand Fragments with variations of parameter values (1) in vertical columns, or request more suggestions for dimension types (2) at the end of the row. Users can further reuse and transfer Fragments to other content via drag-and-drop(3). The textboxes with prompt fragments are shown next to an image, and then a person expands these fragments to show variations. When selecting any of these alternative values of a fragment, the content is modified.}
    \label{fig:fragments}
\end{figure*}

 Fragments reify an initial prompt used to generate text or image into a set of attribute cards, of the format \texttt{\textbf{[type, value]}} (where \texttt{\textbf{type}} is the category of the extracted dimension, and \texttt{\textbf{value}} is the extracted value within that dimension---such as  \texttt{[tone, enchanting]}, \texttt{[content, castle]} or \texttt{[style, illustration]}). 
 Revealing these conceptual dimensions enables an initial reflection-in-intent, revealing the latent structure of the prompt as seen by the AI model. \rev{Commercial software such as Adobe Firefly~\cite{firefly} offers a similar capability as tags, enabling users to select them from a side panel for subsequent image generation. Applying the principle of reification to tags and turning them into cards affords three core novel interactions} as illustrated in Figure~\ref{fig:fragments}.
 
 First, users can reveal fragments via a long press on the content. Fragments are fully reified as interactive instruments and are dynamically generated---hence open-ended and nondeterministic---in contrast to the fixed, hand-crafted, and hard-coded controls supported by prior work (e.g. ~\cite{xia2016object, TerryCreativeNeedsUIDesign2002, Suggestive3dDrawing2007, WritLarge2017}). 
 Second, via drag and drop, users may remove fragments (by dragging them away), or add new fragments onto existing content in the work space. Adding or removing fragments triggers regeneration of the content. 
 Third, to further support reflection, fragments offer suggestions on demand. By tapping on \includegraphics[width=0.015\textwidth]{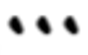}, users can generate new variations from any fragment; these suggestions appear in a column below the specific fragment. Users can also invoke additional suggestions for more types of fragments, which are then appended to the row of fragments. 
 
These three core mechanisms support a workflow where, as users work with multiple images in their workspace, they can explore the effect of different fragments via drag-and-drop to ground one image generation into an aspect of another.

Fragments use the affordance of attribute cards to break down and reify a complex intent into manageable pieces, each having distinct type and value, that enable people to work with these as more-or-less independent and composable, "pieces of intent." This also encourages a workflow where users can surface useful fragmentary concepts surface that become reusable and specialized instruments in their own right. Such fragments are then available for reapplication to other pieces of content, or even reuse in a different context.

While in principle we could have pursued a design that generated many fragments as automatic suggestions associated with each piece of content, such an approach would introduce clutter and risk overwhelming the user with the "decision paralysis" of too many choices. Our design therefore surfaces only a few fragments at a time, and only in a post-hoc manner upon explicit invocation by the user. Further, we present these newly-invoked fragments in an organized fashion, with two orthogonal dimensions of exploration on demand, by keeping dimension type in horizontal rows of cards, and value variations in vertical columns beneath these.

\subsection{Transformative Lenses}

\begin{figure*}[tb]
  \centering
  \includegraphics[width=0.95\textwidth]{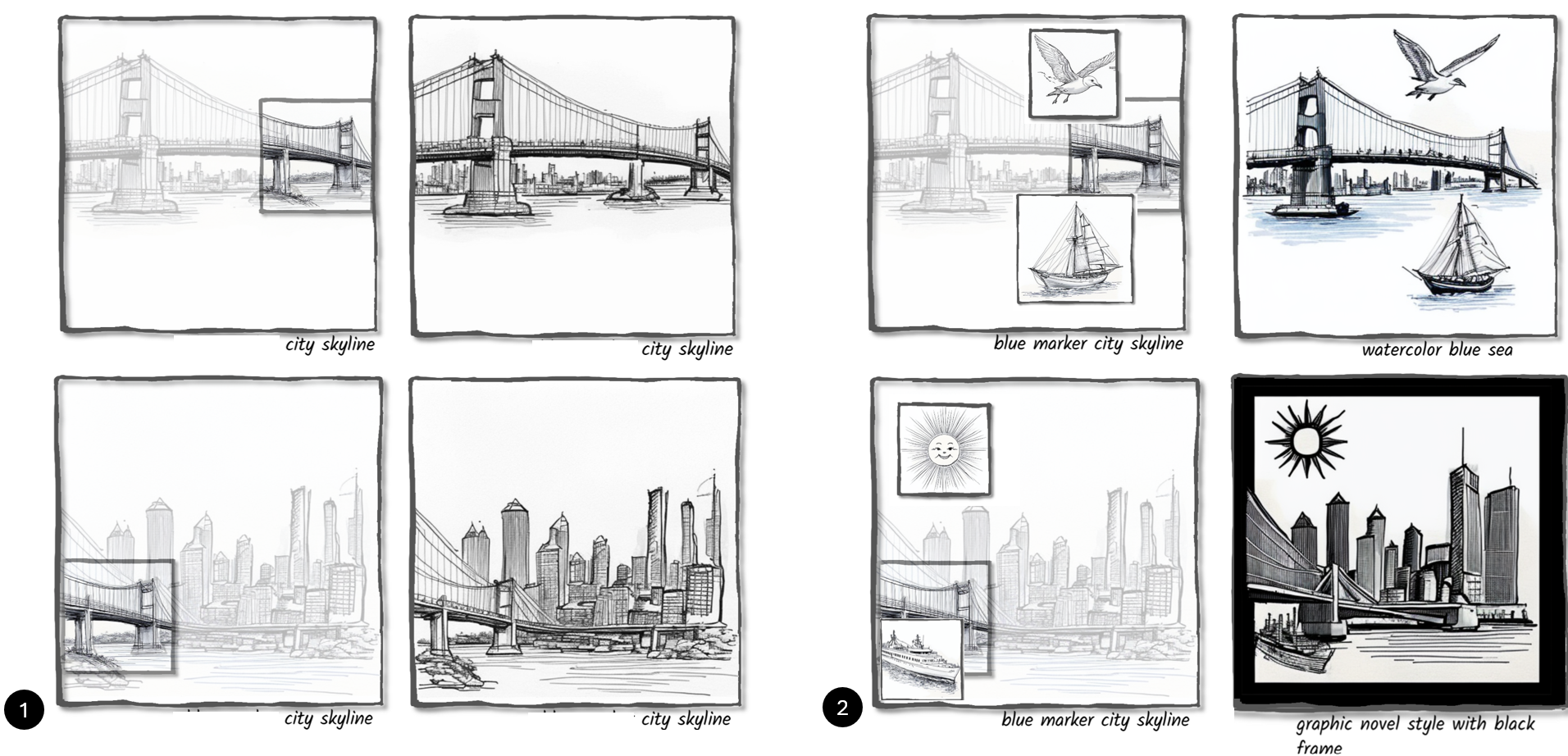}
  
  \caption{Transformative Lenses are placed over initial content, enabling users to "complete" illustrations from pieces of content (1). When users add elements to their composition, lenses regenerate to integrate it (2).}
  \Description{Eight square boxes that illustrate wireframes of the Transformative Lenses. They show in a sequence of how other content gets added to the lens and new images are generated based on the composition of the source images. Transformative Lenses are placed over initial content, enabling users to "complete" illustrations from pieces of content (1). When users add elements to their composition, lenses regenerate to integrate it (2).}
  \label{fig:lenses}
\end{figure*}

Transformative Lenses re-envision the Toolglass and Magic Lens interaction technique ~\cite{bier2023toolglass} as a layered instrument that can be coupled with a generative prompt.

Layering a Transformative Lens on top of content uses such a prompt to generate a new image that synthesizes the lens and the content. Likewise, a specific piece of image content can be used on top of a lens to recombine the two. Such layerings can be positioned and manipulated to chain multiple effects together. 
As illustrated in Figure~\ref{fig:lenses}, users can leverage lenses to take a piece of content (e.g. a sketch of a suspension bridge), and then re-compose this content within a wider backdrop scene (a city skyline), or even apply a new specific style to the results with a single interaction (e.g. a heavy, black-lined graphic novel style). 

More generally, depending on how the user layers Transformative Lenses and image content, lenses can support image completion from a small piece of content, synthesis and composition of multiple pieces of content into a new image, or regeneration of the underlying image. Note also that blank lenses (which have no image content, but do contain a prompt) can be used. For example, a blank-lens backdrop generated afforded outpainting-like operation ---but here steered by the lens's prompt---to "complete" a scene from an existing piece of image content.

Users can freely drag, reposition, and resize both content images and lenses, layering them over each other to chain transformations, reflect on the results, and experiment with different combinations. This property also may encourage users to break down their intent into multiple lenses, which can then be applied to multiple pieces of content (grounding). Note that image recomposition and dynamic regeneration occurs after a 2-second idle time to avoid triggering constant image regenerations during dragging or resizing operations. \rev{As users may wish to adjust content under a lens post-generation, the lens is temporarily faded out in the background when the mouse pointer enters it. }

Complementary to the Fragments described in the previous section, Transformative Lenses afford the design consideration of breaking down the output into pieces (whereas fragments focus on the prompt intent). People can control the composition of images by just moving and layering elements in relation to the lens, limiting the need for precise selection, and encouraging rapid iteration \& experimentation with compositions. However, unlike an undo operation, removing (or otherwise reverting) the layering of Transformative Lens and image-content elements triggers re-generation, and will always lead to a slightly different rendering.

\subsection{Generative Containers}

Designers use moodboards \cite{BuxtonSketchingUserExperiences2007}, storyboards \cite{Storeoboard2016}, and other techniques for presenting small-multiples in  galleries \cite{DesignGalleries1997, TerryCreativeNeedsUIDesign2002, TEAMSTORM2007} to illustrate and explore a space of possible creative directions. \rev{Structured generation of those alternatives \cite{suh_luminate_2024} allows rapid exploration of design spaces, and techniques to highlight similarities and differences \cite{gero_sensemaking_2024} facilitate the selection, refinement, and comparison of multiple responses.}

\begin{figure}[htb]
  \begin{subfigure}{.47\textwidth}
  \centering
    \includegraphics[width=\textwidth]{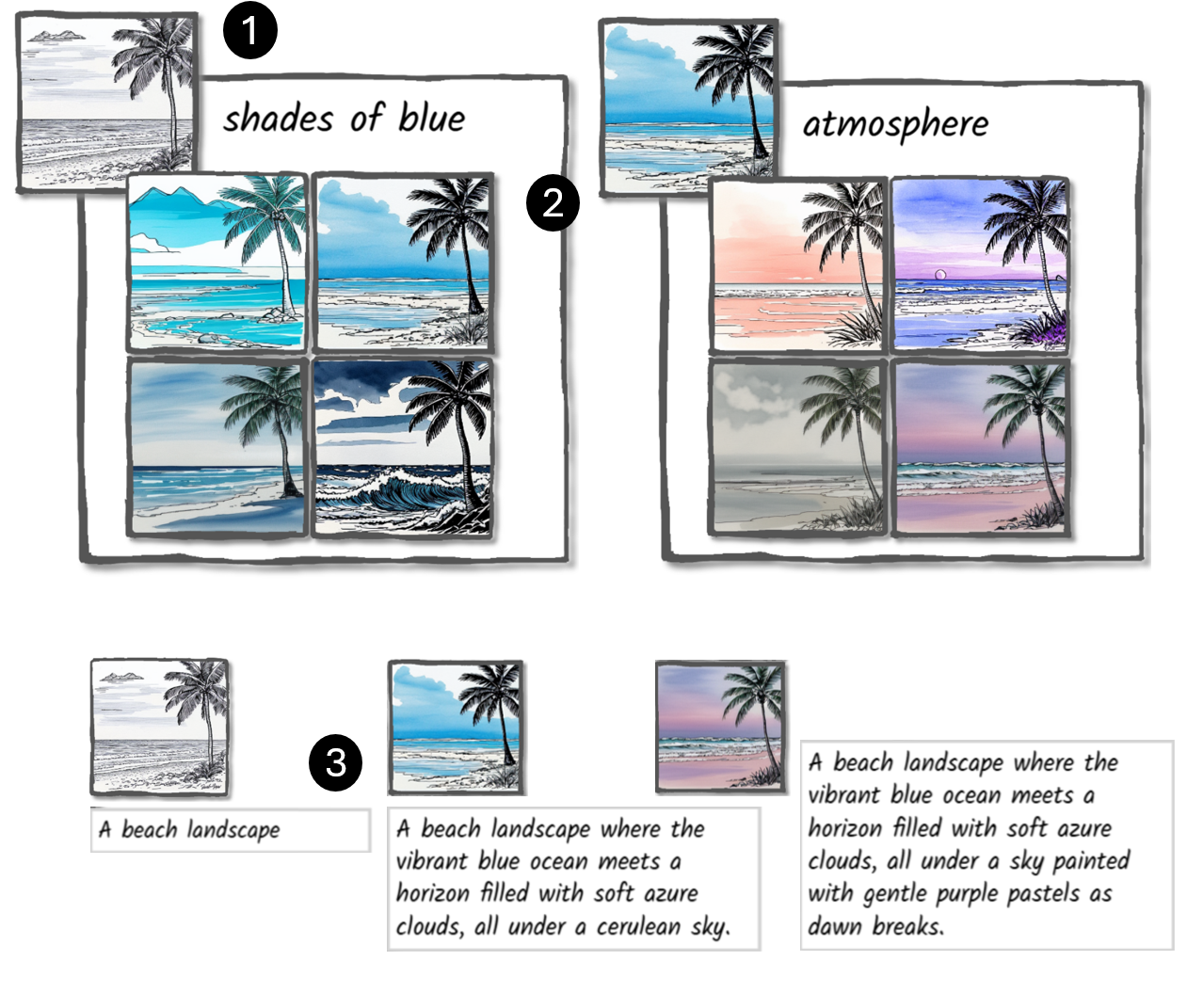}
    \caption{Generative containers enable users to explore possibilities on concrete or abstract dimensions (1). Containers also afford complex exploration paths by reusing the output of one container as the input of another one (2), resulting in refining intent (3).}
    \Description{Generative containers enable users to explore possibilities on concrete or abstract dimensions (1). Containers also afford complex exploration paths by reusing the output of one container as the input of another one (2), resulting in refining intent (3).}
    
    \label{fig:containers}
  \end{subfigure}
  \hfill
  \begin{subfigure}{.47\textwidth}
  \centering
    \vspace{10pt}
    \includegraphics[width=\textwidth]{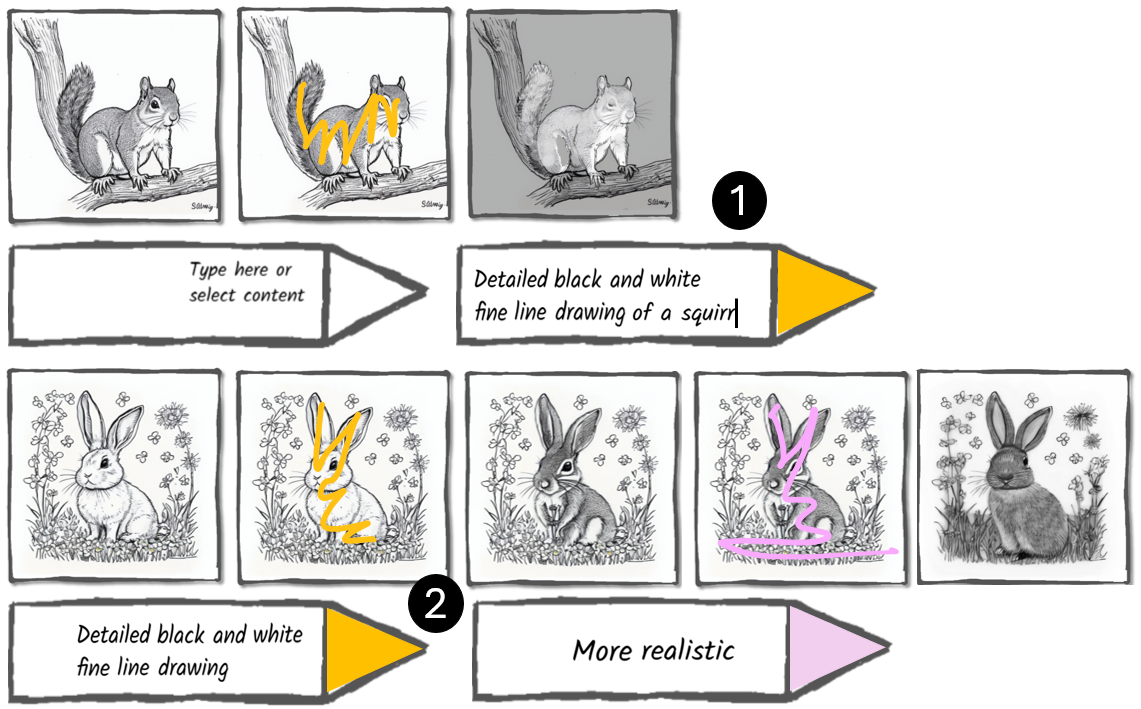}
    \caption{Brushes can extract aspects of content difficult for users to articulate in words, such as drawing style, making it reusable and editable (1). Combined with the selection afforded by brushes, this enables to apply aspects such as style to portions of images (2).}
    \Description{Brushes can extract aspects of content difficult for users to articulate in words, such as drawing style, making it reusable and editable (1). Combined with the selection afforded by brushes, this enables to apply aspects such as style to portions of images (2).}
    \label{fig:brushes}
  \end{subfigure}%
  
  \caption{Generative Containers and Fillable Brushes support different types of content creation tasks. Containers promote the exploration of multiple ideas in parallel, while Brushes offer precise direct manipulation for steering generation. Providing users with both of these AI-instruments enables them to conduct many different activities involved in content creation, enabling  interweaving of both divergent and convergent thinking activities.}
  \Description{Screenshots illustrating generative containers and fillable brushes. Sketched drawings of a beach and palm trees are added to a box that is representing a generative container, which in turn creates variations of those sketches. Generative Containers and Fillable Brushes support different types of content creation tasks. Containers promote the exploration of multiple ideas in parallel, while Brushes offer precise direct manipulation for steering generation. Providing users with both of these AI-instruments enables them to conduct many different activities involved in content creation, enabling  interweaving of both divergent and convergent thinking activities.}
\end{figure}

As shown in Figure~\ref{fig:containers}, \textit{Generative Containers} provide an AI-instrument that encapsulates these notions using a prompt---shown in the container's header---that is closely associated with a 2x2 small-multiple grid of generated image results. Users can then enter or edit the prompt, or drag and drop  example content---or even another instrument, such as a Fragment---onto the Container to ground it and generate a new small-multiple set of results. 

We designed Generative Containers to enable reflection-in-response, allowing users to quickly get a visual sense of the range of responses a single prompt might produce. And by using Generative Containers to generate different variations of fragments, for example to obtain more concrete image editing suggestions from a high-level intent (Figure~\ref{fig:meta-instrument} left), generative containers also enable reflection-in-intent.

In our current implementation, the Generative Containers probe supports generation of four different variations (in a fixed 2x2 grid). Further, each container is presently limited to a single grounding example as input. However, users can create multiple containers and reuse results by dragging and dropping from one to another. 
In this way Containers afford adding details and varying the prompt to generate a range of example images, encouraging multiple cycles of iteration. Recombining and chaining these together effectively results in a longer, refined prompt that integrates the series of changes from prior interactions. 
\rev{Furthermore, one could expand the Generative Container instrument with other representations beyond our 2x2 grid, such as the dimension plots or stacked vertical dimension grids \cite{suh_luminate_2024}.}

\subsection{Fillable Brushes}

Fillable Brushes, as illustrated in Figure~\ref{fig:brushes}, 
offer an AI-instrument with the semantics of an "intelligent paint brush" for style transfer scoped to a particular spot on an existing image. 

While previous work has explored brushes that can encapsulate and integrate deterministic modes and commands ~\cite{romat2022style}, our Fillable Brushes instrument applies encapsulated AI-prompts onto content in an intelligent manner as the user scrubs over it with their pen, finger, or other pointing device. And in contrast to the post-hoc notion of Fragments described above, Fillable Brushes apply a brush onto content, structured as an AI-Instrument "command" with a prefix (as opposed to postfix) syntax ~\cite{LexicalPragmaticInput1983}. This offers a familiar interaction model from the way that a highlighting tool turns selected text yellow in a document editor, for example.

\rev{Encapsulating a prompt or image into a brush to define its function is a powerful interaction techniques to control scope of selection, as demonstrated by Runway motion brushes~\cite{runwayML}. Applying additional principles of our model, enables users to} also "fill" (ground) an empty brush by using existing content as an example, as if the instrument were a color picker that picks up key semantic attributes of the content rather than just its "color." The prompt encapsulated by the Fillable Brush is then automatically populated with descriptive words via generative AI, which the user can further edit if desired. This can be particularly helpful when users want to style something "like this" even when they may lack the vocabulary to describe its visual style. Our Fillable Brushes technology probe supports both content and/or style extraction. \rev{Turning a brush into a persistent object on screen, enables combining brushes together by drag and drop.} Brushes can also be applied multiple times to the same content to emphasize a particular prompt in the result. 

While Fillable Brushes enable users to specify the scope of intent with a high granularity, this does not necessarily require high precision: our implementation leverages the AI-powered Segment Anything Model (SAM)~\cite{kirillov_segment_2023}, which enables users to make approximate selections (i.e. rather than a precise and tedious lasso selection) to indicate an image element. The source content plus the approximate selection (as a set of reference points) is then converted into a precise object selection by the segmentation model.

\subsection{Generated Instruments and Meta-Instruments}

\begin{figure*}
    \centering
    \includegraphics[width=\linewidth]{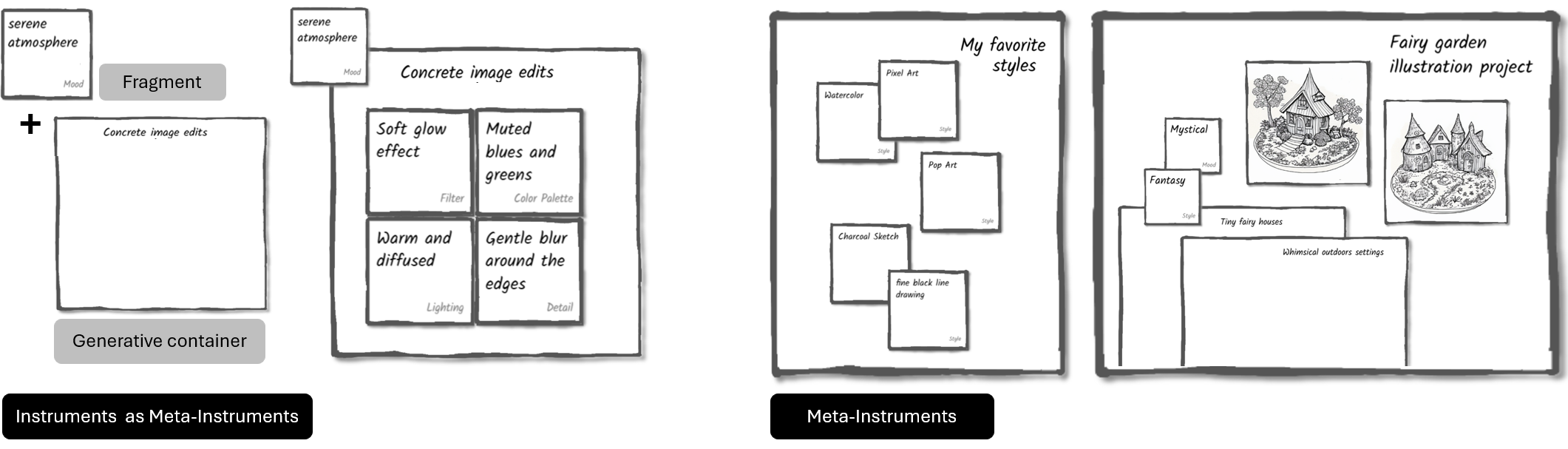}
    \caption{Instruments can be used as Meta-Instrument: operate on each other to create related instruments, for example for making a fragment more concrete (left). Specific Meta-Instruments such as palettes (right) can help user organize sets of instruments for easier retrieval and reuse, or, even help generating collection of instruments for a certain task.}
    \Description{Wireframe and screenshots (multiple squares that represent the composition of multiple AI instrument in a meta instrument) -- Instruments can be used as Meta-Instrument: operate on each other to create related instruments, for example for making a fragment more concrete (left). Specific Meta-Instruments such as palettes (right) can help user organize sets of instruments for easier retrieval and reuse, or, even help generating collection of instruments for a certain task.}
    \label{fig:meta-instrument}
\end{figure*}

Beyond the concept of instruments, the instrumental interaction model~\cite{beaudouin2000instrumental} also refers to the concept of meta-instruments, in which \textit{"instruments operate on instruments"}. 
As hinted at in earlier sections, using instruments on other instruments can be particularly useful to derive or compose instruments from the "task detritus" \cite{KirshIntelligentUseOfSpace1995} already produced in the user's workflow and experimentation with other instruments. Using generative containers on Fragments, for example, can help users navigate the degree of abstraction, turning a vague idea into a set of concrete modifications (Figure~\ref{fig:meta-instrument} left). 

However, such generation loops (instruments that generate content, generating instruments that generate other instruments generating content...) could potentially lead to an unwieldy number of elements in the interface. To organize but also generate collections of instruments, we devised a type of meta-instrument we call \textbf{Palettes}.

Akin to menus and containers available in GUIs today, Palettes enable storage and/or generation of different sets of instruments and content if desired. These afford abstraction and generalization of instrumental controls from collected pieces of content that can then serve as examples or generative seeds (Figure~\ref{fig:meta-instrument} right). Palettes of diverse instruments can balance the different affordances and properties of each instrument to provide rich content creation support. They can also help people get past the "cold-start" problem in complex creative design work, by beginning with examples, other pieces of existing content, or past work-artifacts to help overcome so-called "writer's block" or "blank canvas" effects of starting from nothing.

\section{Implementation}

\paragraph{Overview:}
Our system and all AI-Instruments technology probes were implemented on a web-based platform. We use Javascript and HTML with the fabric.js~\cite{juriy_zaytsev_fabricjs_2024} library for the front-end, and a Node.js~\cite{openjs_foundation_nodejs_2024} server for the back-end managing content and files as well as coordinating communication  with the generative AI models. For the user interface design, we chose to use a sketched user interface look and feel, to encourage our study participants to focus on the concepts rather than the surface details of their specific instantiation in the UI \cite{BuxtonSketchingUserExperiences2007}.  

\paragraph{Leveraging Generative AI models:}
We use the OpenAI GPT-4o~\cite{openai_gpt-4o_2024} model for text transformations and image analysis, and a local Stable Diffusion~\cite{stabilityai_compvisstable-diffusion_nodate} server with a custom processing pipeline for image generation. The AI-Instruments use GPT-4o for analysis of provided input (e.g., for turning provided visual content into a text prompt, analyzing the contents of part of the workspace canvas). For image generation, we use multiple stacked ControlNet~\cite{zhang_controlnet_2023} models with Stable Diffusion to steer the generation of visual content. To preserve aspects of the source input, we use a combination of \textit{Depth}, \textit{Canny Edge}, and \textit{Scribble} ControlNet models, while for preserving art/rendering styles (e.g., the sketch-based output) we use the \textit{Reference} ControlNet model. Depending on the type of image generation, we vary the weight of each ControlNet model (e.g., increasing weight to emphasize content preservation, or decrease weight of another ControlNet to reduce affect of reference style transfer). We use image masks to selectively control which areas are changed or kept, apply inpainting/outpainting scripts, and adjust other parameters such as CFG scale, denoising strength, and control mode. 

\paragraph{Building AI-Instruments:}
We designed a pipeline that can orchestrate the access and requests to the different LLM and diffusion server instances to generate results. Key functionality is wrapped in modules, such as for encapsulating prompts to communicate with one or more models (by using model chaining) to perform a specific task. 
Each AI-instrument then uses a number of these modules for modifying the input or generating new content bases on the user's performed action with the instrument:

\begin{itemize}
    
\item \textit{\textbf{Fragments}} instruments include modules for (1) prompt decomposition which takes a text or visual input and makes a GPT-4o request to generate fragments (returned as collections of \texttt{[type, value]} pairs), (2) fragment extension which takes a prompt and the existing fragments and requests additional fragment dimensions, (3) fragment variation which takes the fragment and parent prompt/content (if applicable) and generates variations of that fragment, and (4) prompt composition which takes a prompt, a modification to the fragments, and returns a modified prompt. The result from the prompt composition is then used to create an updated image with the Stable Diffusion + ControlNet pipeline.

\item \textit{\textbf{Transformative Lenses}} use a module for composition of the prompt (merging prompts from all source images covered by the lens), before then applying inpainting/outpainting, masks, and ControlNet models to generate the resulting image.

\item \textit{\textbf{Containers}} use a variation module, taking a prompt and a dimension, and requesting four variations along the provided dimension. Within the prompt we request visually diverse results. The resulting set of prompts is then sent to SD+ControlNet to generate the final set of four images in the container.

\item \textbf{\textit{Fillable Brushes}} are implemented to either emphasize style or content variations, depending on the intent of the user, which we support by varying the weight of the ControlNet models (e.g., higher reference ControlNet weight for changing the visual style, or increasing weight of Canny-edge/Depth ControlNet to preserve existing content).

When the brush is applied, we perform a segmentation of the source content by feeding the stroke path as control points into Segment Anything~\cite{kirillov_segment_2023}, which results in a segmentation mask of the dominant object(s) selected with the brush stroke. We then use GPT-4o to craft a combined prompt given the source image(s), the segmented content, and the original prompt. Finally, we send this generated prompt together with the source image and segmentation mask to the Stable Diffusion server, using the ControlNet inpainting method.

\end{itemize}
 
\section{Study}

We conducted a qualitative user study with 12 participants to gather \rev{their insights on AI-instruments compared to traditional prompting}. Participants completed image generation and editing tasks with our four technology probes as well as an initial chat-based prompting probe, to help them tease out pros and cons of these two different interaction models. We analyzed their comments to understand the perception of the principles of reification, reflection and grounding, as well as capture insights on the Human-AI interaction challenges (listed in Section ~\ref{sec:challenges}) addressed by different instruments.

\subsection{Procedure}
Participants completed a 60-minute study in a quiet room on a computer running our technological probes and a study form. After obtaining informed consent, we collected basic demographic information, then requested participants to complete a set of tasks with our technological probes. All participants first completed tasks with a chat-based prompting probe using the same image generation model as the instruments in order for us to confirm their familiarity with prompting and to also provide them with a baseline for the image generation model with use. Before each instrument, participants watched a video demonstration explaining functionalities and modalities of interaction. After this video, participants used the technological probes to complete 2 to 3 tasks such as generating an image and changing its style. We provided example content for each task, but encouraged participants to generate their own content and try different ideas. The experimenter only interacted with participants during this phase of the study to clarify functionalities and interaction if needed. After each set of tasks, participants reflected on one key advantage and on key inconvenient of this specific instrument compared with the chat-based prompting interfaces. Since we aimed at gathering qualitative insights on the overall interaction model behind instruments rather than compare instruments against each other, all participants completed the tasks in the same order. After tasks completion, we described five types of generic tasks, asking for each to select the best interaction technique. Participants also entered the rationale for their choice. At the end of the study, participants received a \$50 gratuity for their time. The study protocol was reviewed and approved by the Microsoft Research ethics review board.

\subsection{Participants}
We recruited 12 participants (8 men, 3 women, 1 non-binary) via mailing lists in a large organization. We \rev{selected participants with weekly interaction with} generative AI systems (ChatGPT, stable diffusion, etc) for creating content. \rev{As we aimed at gathering insights on overarching interaction principles for AI-instruments (across many domains), we opted for selecting participants with interesting in different types of content generation. Our participant pool included users interested in authoring short text snippets such as emails, long structured documents such as reports, structured text such as tables, programming code and web-pages, visual artifacts such as images, and multimodal artifacts with text image and charts such as presentations. Note that none of our participants generated audio or video.} 

\subsection{Material and Analysis}

We collected the salient advantage and salient weakness for each instrument compared to prompting. Participants experienced the following probes: 1) chat-based prompting, 2) fragments, 3) containers, 4) lenses, and 5) brushes. To encourage participants to think of different aspects of content generation, we asked them their preferred interaction technique (along with their rationale) for five different tasks. 

\begin{enumerate}[leftmargin=*]
    \item [\textbf{T1}] \textbf{Combining content:} merging pieces of content together
    \item [\textbf{T2}] \textbf{Splitting content:} extracting a piece of content
    \item [\textbf{T3}] \textbf{Iterating on content:} editing an aspect of content  
    \item [\textbf{T4}] \textbf{Editing by example:} transferring content or style
    \item [\textbf{T5}] \textbf{Expanding content:} adding new material to existing content
\end{enumerate}

We coded a total of 156 statements from our participants to gain insights on their perception of our model's principles and to assess how AI-instruments (un)successfully addressed the five challenges \rev{described in section~\ref{sec:challenges}}. A portion of these comments (28/156) also revealed limitations of our technical probes (the codebook is available at \url{https://hugoromat.github.io/ai_instruments/}).

\subsection{Insights on Model Principles}

\paragraph{\textbf{Reification of intent}} All 12 participants reacted positively to the principle of reifying intent into AI-instruments. Participants valued that AI-instruments enabled them to shift the focus from the prompt to the outcome. P6 commented that \textit{"I can just click on the fragments instead of typing it out and focus on the final output instead."}. P10 noted it was helping them with the iterative process: \textit{"with prompting it makes me think of the prompt, but having [fragments] already in front of me can make it easier for me to make a decision of what i want."}. They also valued the \textbf{direct interaction} afforded by AI-instruments: \textit{"[with fillable brushes] you can interact with the images that are generated directly, rather than [modifying] the image only from prompting."}

A few participants also outlined the value of reification for storing and reusing prompts. P9 commented on Fillable Brushes \textit{"[...] I would be able to create a [brush] with the thing that I wanted to pull out and then apply/store it however I wanted."}, and P11 on Transformative Lenses \textit{"I like the idea of creating and saving a lens and applying it consistently to different images for future uses"}. 

All 12 participants also commented on \textbf{scope of selection} as a key advantage of AI-instruments over prompting (38 comments). Participants identified Fillable Brushes as enabling them to specify portion of an image while Transformative Lenses enabled them to combine multiple images together.  With Brushes, participants emphasize the granularity of the selection. For example P1 noted \textit{"I can highlight the part of the image that requires changing only, and it seems I can highlight at a very high granular level, such as a face of a dog."} P7 referred to this capacity for steering image generation as one can use masks in graphics editors \textit{"I can control a more fine grained area/mask that I want to edit. It's so cool! "}  P4 noted Lenses were particularly useful for adding elements iteratively: \textit{"about how to incrementally add new items from an initial picture, the others [AI-instruments] are more for customize different picture styles"}. In addition, participants appreciated controlling image composition with Lenses, as P2 explained: \textit{"Being able to specify the location of component objects is really helpful"}.

\paragraph{\textbf{Reflection}}

All 12 participants pointed to \textbf{reflection-in-intent} as an advantage of AI-instruments in contrast to prompting (31 comments). This principle was particularly highlighted as a strength of Fragments (24/31 comments), as P5 explained \textit{"I could tell what the different aspects of the prompts were being split [...] and guess what the AI interpreted as something other than what I had in mind."} Participants also praised the benefit of generating variations for each fragment such as P11: \textit{"the system generates ideas for you which you can implement, allowing you to add variables which you might not have originally thought of."}

All 12 participants also outlined \textbf{reflection-in-response} as an advantage of AI-instruments in contrast to prompting (31 comments). Generative Containers were mostly (26/31) cited for this ability. P5 explained that \textit{"Usually I ask AI to give me a different version/example [of text], but this would allow you to choose the one you like without needing to prompt it again."} P2 valued this ability for iterating: \textit{"[generative containers are a] great abstraction for deciding what to iterate on given multiple possibilities. Makes it easier to visualize or use results from previous steps"}.

\paragraph{\textbf{Grounding}}

All 12 participants identified grounding as an advantage of AI-instruments over prompting (45 comments). It was especially noted valuable for images, as P4 explains \textit{"apply styles of different pictures into other ones, sometimes the styles are hard to illustrate in prompting, since they are more abstract"}.

A majority of participants referred to grounding as a key advantage of Fillable Brushes. P8 referred to grounding for dealing with multiple items: \textit{"[...] having the flexibility to copy any kind of prompt on the brush is good if I have similar kind of images to replicate."} P9 referred to grounding for iterating on a single item: \textit{"[brushes] can give us a style for A (which we can store and ensure that it gets applied to all our later additions.)"}.

\paragraph{\textbf{Model limitations}} 

We gathered 20/156 comments pertaining to the model limitations.

Four participants described \textbf{limitations of GUIs} \rev{compared to} chat-based conversational interactions with AI. For example, P11 described accessibility issues of relying on 2D interactions as opposed to the ability to rely solely on speech as is easily feasible with chat-based prompting interaction. P2 mentioned AI-Instruments would suffer from the same issues they encountered with GUIs: ~\textit{"GUIs in general are more prone to bugs or unexpected behaviors than text interactions, which could lead to increased computation that the user doesn't actually want"}. And P10 noted that they felt they needed to perform many interactions ~\textit{"compared to prompting, where I can just type something"}, P12 concluding that \textit{"It's faster to get exactly what I want with prompting"}. These comments do not address fundamental limitations of the instrumental model per say but rather highlight that users may sometimes prefer a single albeit limited modality of interaction. These insights hints at the need to integrate chat-based prompting interaction more tightly with instruments. \rev{A straightforward avenue for this, is to provide on-demand access to underlying textual prompts generated by instruments}.

P6 noted that \textit{"It might be better to write a new prompt when I need to make major modifications to the output"} as a drawback of Lenses. This echoes the sentiment of three other participants in that the grounding capabilities of AI-instruments such as Lenses and Brushes enable effective content generation steering but may hinder divergent content generation. However, this is also balanced by the strengths of other AI-instruments such as Generative Containers that many participants praised for \textit{"creative and exploratory settings."} (P12). These insights hints at the need to \textbf{provide multiple instruments} to users in content generation tools. 

We devised our four AI-instruments to cover different facets and tasks involved in  content creation. For example, Containers supports iterative exploration by reusing pieces of content in other Containers, while Fragments enables it by selecting different aspects to vary. The different design decisions tied to the affordances of each instrument led to different task support, especially with regard to navigating creation history.  This caused occasional frustration for Fragments or Lenses \textit{"Old edits get lost when new edited are prompted"} (P7). Two participants reflected on this strength of chat-based prompting interfaces to inherently provide a trace of all the prompts and results made in a temporal manner. Especially in the context of non-deterministic outputs, where the next piece of content may be less satisfactory than the prior one, these insights suggest that more thoughts need to be devoted to \textbf{systematically surface history} within each instrument.

Note that a majority of aspects participants did not value for AI-instruments (28/48) pertained to the limitations of our technology probes rather than limitations of the model. Many of these comments referred to usability issues such as the fact that Lenses regenerated too early (or too late) or exclusively had a square aspect ratio, as well as visual look and feel of the probes \textit{"color the pens maybe? easier to organize and distinguish them"}(P7).

\subsection{Challenges Addressed by AI-instruments}

\paragraph{\textbf{C1 Intent formulation}}

Participants highlighted \textbf{grounding} to address intent formulation on images, especially as instantiated in Fillable Brushes for extracting and transferring styles.  P7 summarized how Brushes help formulate intent with both scope of selection and direct manipulation: \textit{"I can control a more fine grand area/mask that I want to edit. It's so cool! I can also copy the style/content from a different image and directly apply it to a different image without having to find the accurate words to describe them."}

\paragraph{\textbf{C2 Intent disambiguation}}

Participants identified \textbf{grounding and reflection} as core principles helping them disambiguate their intent.  They explained that the reflection-in-intent surfaced in Fragments was helping them identify context and details to refine their intent. P5 also noted that Fragments enabled to experiment and \rev{gain an understanding} of how the model worked \textit{"allow[ing] me to see how the prompts were being isolated and used"}. A few participants also explained that Brushes helped with intent disambiguation by capturing one by example and expressing it words. P1 summarized: \textit{"one advantage [of brushes] is that it can identify a style, even though I cannot articulate the style well, this is especially helpful to circumvent prompt engineering."}

\paragraph{\textbf{C3 Intent resolution}}

Participants outlined the principles of \textbf{reification and reflection} as most useful for resolving intent. Participants explained that reification enabled them to explore different paths by iterating on different images: \textit{"this lets you build off of previous iterations and you can create different styles for the same image until you are happy with one"}. They highlighted the benefit of reflection-in-response of Generative Containers to help them explore possibilities:\textit{"I don't have to think too much about what I want, which is helpful"} (P10), \textit{"Very easy to explore creative options, it helps to get a better idea of what you like"} (P12). P7 valued the ability to start from high-level prompts and follow up by making a selection~\textit{"It it easy to generate images with a vague description, and offer options [to explore]"}. A few participants also mentioned the benefit of reflection-in-intent of Fragments to experiment with options: \textit{"[Fragment] also suggests some dimensions to change the picture which I might have not thought about. Like the elephant in this case."} (P9). 

\paragraph{\textbf{C4 Steering}}

All participants mentioned the benefit of instruments over chat-based prompting for steering content. Participants noted that direct manipulation and scope of selection afforded by \textbf{reification} were critical in editing portions of content, in conjunction with \textbf{grounding} to capture and transfer aspects of one content to another. Many participants outlined the value of Brushes and Lenses to generate masks for steering content generation \textit{"masking [with lenses] allows me to personalize smaller parts of the image compared to generating a completely new one"} (P10), \textit{"this [brush] is an advanced way to mask out single things within an image to regenerate"} (P3).  P6 summarized the benefits of these instruments compare to chat-based prompting: \textit{"[with brushes] I can quickly make modifications to the image instead of writing a prompt from scratch for every modification. I can focus on the subject I'm interested in."}

\paragraph{\textbf{C5 Workflows}}

Participants perceived that AI-instruments supported two workflows that were currently hard to achieve with chat-based prompting interaction: \textbf{non-linear exploration} and \textbf{iterative content generation} by chaining prompts together (e.g. steering). For exploration, participants each favored different instruments. P6 favored Fragments \textit{"I feel this [fragments] makes it easy to try out different examples that I might be interested in. I can just click on the fragments instead of typing it out and focus on the final output instead."}. P3 described Containers allowing \textit{"more freedom to explore more options. Downside to prompting is that you need to know what you want."} P8 valued \textit{"trying different lenses on the same component creating a scene is easy way to experiment and merge different style in one"}. P8 also referred to the grounding capabilities of Brushes as useful to experiment with the collection of examples offered \textit{"different kinds of image styles to choose from for copying styles or format"}. For iterative content generation, participants refer to the same advantages AI-instruments offer than for steering.

\section{Discussion and Future Work}

We first discuss how our model revisits and extends the classic instrumental interaction model, then discuss the application of AI-instruments to different forms of content, outlining future work.

\subsection{Revisiting and Extending the Instrumental Interaction Model}

The instrumental interaction model~\cite{beaudouin2000instrumental} is based on three core principles~\cite{beaudouin2000reification}:
\begin{itemize}
    \item \textit{Reification of commands} refers to the principle of turning systems functionalities into interactive graphical objects in the interface,
    \item \textit{Polymorphism} refers to the principle of applying commands to different types of objects enabling the interface designers to keep the number of interface objects relatively small, 
    \item \textit{Reuse} refers to the ability for users to reapply one command to different objects or apply different commands to one object, with the goal of limiting repetitive user input and/or navigation.
\end{itemize}

In this paper, we revisited and proposed to extend this model in the following ways:

(1) We extended the principle of reification from encompassing a limited set of commands defined for an application to include any intent user may express in natural language. We also unpack two key considerations of reification that one should consider when designing AI-Instruments: the scope of the instrument, and degree of abstraction. We propose to leverage the affordances of existing direct manipulation techniques to convey to users how to specify scope (e.g. select a portion with a brush vs resize the lens). To support users navigating different levels of abstraction of their intent, we propose to use AI-instruments themselves.

(2) We re-framed the concept of polymorphism in instrumental interaction, recasting it as \textit{reflection}. This shifts instrumental interaction from a classic "direct manipulation" technique for graphical user interfaces to a modern AI-augmented technique. Reflection leverages the general concept-translation capabilities of LLM's to support an expansive notion of "polymorphism" without requiring the interface design and system architect to hand-code the parameters, controls, and nuances of how these are interpreted across a wide range of content types. 

Further, by considering reflection from both the user (intent) and system (response) perspectives, we provide users with mechanism to explore both the design space of their intent (i.e. different formulations and disambiguation of intent), as well as the design space of the model response (i.e. different interpretations of user intent by the model). While such notions, in one sense, have been latent in interactive instruments all along, with generative AI many possible forms and interpretations of polymorphism---potentially even for niche or specialized workflows, formats, and types of content, if they are sufficiently represented in the training data for the model---can be made available for user reflection in AI-assisted content creation.

(3) The third principle of reuse is closely related to our principle of grounding. Grounding extends the notion of \textit{reusing commands} to the capability of \textit{extracting and reusing intent}---whether in terms of one aspect of user intent, a collection of multiple user intents, or other properties of content. This principle of grounding also encapsulates the ability to generate instruments from other instruments, characterized as \textit{meta-instruments} in the nomenclature of instrumental interaction \cite{beaudouin2000instrumental, beaudouin2000reification}.

(4) A further new challenge raised by AI-instruments is the need to balance the possibility of over-generating instruments, with the power to encapsulate many capabilities---at a high level of abstraction---within a single instrument. In contrast to classic hand-crafted instrumental interaction, AI-generated instruments could potentially lead to an unwieldy number of objects in the workspace, if generation were left unchecked. However, we counterbalance this with strategies to compose instruments and organize them into collections (meta-instruments). Further, we can leverage the generative nature of AI to iteratively refine both content and (meta-)instruments---altering, summarizing, or abstracting instruments and content with each step---as another strategy to harness AI to express aggregated concepts at a high semantic level.

The insights we gained by building a set of technology probes and gathering initial perceptions of 12 content creators, suggest that the principles described in our model can be used inform the design of novel interaction techniques as well as assess existing ones. \rev{As we built each probe, it became clear that design decisions at lower-level, for example pertaining to the specific choice of interactions (e.g. click vs double-click) or their timing (e.g. idle time threshold triggering image generation), can lead to different experiences, especially when multiple probes are used in conjunction.  While our model suggests overarching principles for AI-instruments, specific interaction is bound to differ as sets of these techniques are integrated into specific applications and adapted to different modalities and contexts~\cite{appert2005context, mackay2002interaction}.}

\rev{Additional considerations for integrating AI-instruments in applications include the expectations of users with regard to direct manipulation and instrumental interfaces, as well as when working with AI. For example, a fundamental principle of direct manipulation is ease of reversibility of user actions (e.g. if a fragment is removed from an image triggering a new generation, then added back; the image should revert to its prior state). In contrast, AI models are non-deterministic by nature (e.g. same input, different output). While it is technically feasible to couple AI generation with history and versioning mechanisms to ensure the reversibility of operations, users' attitude on working with AI over the longer term, as well as specific use cases may impact this design decision. }

\subsection{Beyond Images, Applying AI-Instruments to Other Forms of Content}

While extrapolations from participants must be taken with caution, four participants in our study related the use of AI-instrument to content they worked with every day. Interestingly two participants (P2 and P9) commented they would not see the use of instruments for tasks such as writing code. P9 summarized it as \textit{"it [fragments] is a bit harder to use than prompting for tasks like maybe writing code. I would prefer fragmenting for images or plots"}.  However, two different participants (P11 and P12) envisioned using AI-instruments for data analysis and writing (P11) and in the case of P12 leveraging grounding for operating at the artifact level: \textit{"it would be so cool to auto-pick a style (words, design, etc) without me first having to decipher it, and then have it automatically apply to other content (writing, slides, etc)"}.

We experimented with a few of our technology probes (Fragments, Containers and Palettes) to work with textual content and found that our principles generally held. However, further exploration with different types of modality is likely to reveal additional design considerations for the instruments we proposed. Notably, a key issue to address for textual content as opposed to images is the effort required to consume a number of potential outputs (reflection-in-response). Integrating support for helping user skim and get the gist of similarities and differences between textual outputs in Generative Containers (by bolding portions of text or providing summaries or excerpts) would certainly be necessary when using instruments to work with textual documents.  

Another aspect to address is the use of instruments for artifacts composed of multiple pieces of content (e.g. a slide composed of a title and image). Again, while we believe core principles hold for devising instruments to work with content at the artifact level, additional research is needed to delve into how to integrate different aspects of an artifact. For example, one could envision displaying Fragments from different scope of selection, enabling Fragments to operate at the entire slide level or on a subset such as title.

In the future, we plan to pursue these two research directions (designing AI-instruments for heterogeneous content and artifacts composed of multiple pieces of content), further assessing the generalizability of our interaction model and broadening the set of design considerations for AI-instruments.

\section{Conclusion}
We operationalized the theory of instrumental interaction for generative AI, with an in-depth unpacking of the principles of reification of user intent, reflection, and grounding. We argue that leveraging this re-appropriated and refined theory can drive the creation of a \textit{new generation of expressive AI-Instruments} that afford better expression of intent, make it easier to discover what is possible, and provide powerful degrees of freedom for steering the generation towards the best possible results. Those new tools and instruments can truly leverage the polymorphic and non-deterministic behavior of generative AI models, unleashing new and empowering forms of expressive HCI+AI experiences. 

Beyond our focus on AI-Instruments, theories play an important role in the advancement of our wider research field~\cite{rogers_hci_2012, halverson_activity_2002}. Rogers argues that there is a need for theories as lenses bringing critical design characteristics into focus, and which can function as a generative source: providing "\textit{design dimensions and constructs to inform the design and selection of interactive representations}"~\cite{rogers_new_2004}. We hope that our work on operationalizing the theory of instrumental interaction for AI can inspire other new -- and re-appropriated -- theories to advance HCI+AI.

\begin{acks}

We thank the participants of our AI-Instruments user study for taking the time to provide feedback on our techniques, and the reviewers of this submission for their constructive suggestions to improve this research.

\end{acks}

\bibliographystyle{ACM-Reference-Format}
\bibliography{References}

\onecolumn
\appendix

\end{document}